\documentclass[preprint,floatfix,noshowpacs]{revtex4}%
\usepackage{amsfonts}
\usepackage{amsmath}
\usepackage{amssymb}
\usepackage{slashed}
\usepackage{graphicx}
\def\Phimq{\Phi_M^{(q)}}

\def\phimqi{\phi_M^{(q_i)}}
\def\phimqq0{\phi_M^{(q_0)}}
\def\phimq8{\phi_M^{(q_8)}}
\def\phimql{\phi_M^{(q_\ell)}}
\def\phimqs{\phi_M^{(q_s)}}

\def\phimg{\phi_M^{(g)}}

\def\fmi{f_{M}^{i}}

\def\fmq0{f_{M}^{0}}
\def\fm8{f_{M}^{8}}
\def\fml{f_{M}^{\ell}}
\def\fms{f_{M}^{s}}

\def\aqimn{a^{(q_i)}_{Mn}}
\def\agmn{a^{(g)}_{Mn}}

\def\aqqq0mn{a^{(q_0)}_{Mn}}
\def\abqqq0mn{\tilde{a}_{Mn}^{(q_0)}}
\def\aqq8mn{a^{(q_8)}_{Mn}}
\def\aq8mk{a^{(q_8)}_{Mk}}
\def\aqsmn{a^{(q_s)}_{Mn}}
\def\aqlmn{a^{(q_\ell)}_{Mn}}

\def\aetnq0{a_{\eta n}^{(q_0)}}
\def\aetnqq8{a_{\eta n}^{(q_8)}}
\def\aetng{a_{\eta n}^{(g)}}

\def\aetpnq0{a_{\eta^{\prime}n}^{(q_0)}}
\def\aetpnqq8{a_{\eta^{\prime}n}^{(q_8)}}
\def\aetpng{a_{\eta^{\prime}n}^{(g)}}

\def\aetnql{a_{\eta n}^{(q_l)}}
\def\aetnqs{a_{\eta n}^{(q_s)}}

\def\aetpnql{a_{\eta^{\prime}n}^{(q_l)}}
\def\aetpnqs{a_{\eta^{\prime}n}^{(q_s)}}

\def\aplmn{a_{Mn}^{\left(  +\right)  }}
\def\amimn{a_{Mn}^{\left(  -\right)  }}

\def\phietl{\phi^{(q_\ell)}_\eta}
\def\phiets{\phi^{(q_s)}_\eta}
\def\phietpl{\phi^{(q_\ell)}_{\eta^{\prime}}}
\def\phietps{\phi^{(q_s)}_{\eta^{\prime}}}

\def\Fmg{F_{M\gamma}}

\begin{document}
\title{$\eta$-$\gamma$ and $\eta^{\prime}$-$\gamma$ transition form factors
in a nonlocal NJL model}
\author{D. G\'{o}mez Dumm$^{a,b}$, S. Noguera$^{c}$ and N.N. Scoccola$^{b,d,e}$}
\affiliation{$^{a}$ IFLP, CONICET $-$ Departamento de F\'{\i}sica, Facultad
de Ciencias Exactas, Universidad Nacional de La Plata, C.C. 67, 1900 La
Plata, Argentina}
\affiliation{$^{b}$ CONICET, Rivadavia 1917, 1033 Buenos
Aires, Argentina}
\affiliation{$^{c}$ Departamento de F\'{\i}sica
Te\'{o}rica and IFIC, Centro Mixto Universidad de Valencia-CSIC, E-46100
Burjassot (Valencia), Spain}
\affiliation{$^{d}$ Physics Department,
Comisi\'{o}n Nacional de Energ\'{\i}a At\'{o}mica, }
\affiliation{Avenida del Libertador
Libertador 8250, 1429 Buenos Aires, Argentina}
\affiliation{$^{e}$
Universidad Favaloro, Sol{\'{\i}}s 453, 1078 Buenos Aires, Argentina}

\begin{abstract}
We study the $\eta$ and $\eta^{\prime}$ distribution amplitudes (DAs) in the
context of a nonlocal ${\rm SU(3)}_{L}\otimes {\rm SU(3)}_R$ chiral quark
model. The corresponding Lagrangian allows to reproduce the phenomenological
values of pseudoscalar meson masses and decay constants, as well as the
momentum dependence of the quark propagator arising from lattice
calculations. It is found that the obtained DAs have two symmetric maxima,
which arise from new contributions generated by the nonlocal character of
the interactions. These DAs are then applied to the calculation of the
$\eta$-$\gamma$ and $\eta^{\prime}$-$\gamma$ transition form factors.
Implications of our results regarding higher twist corrections and/or
contributions to the transition form factors originated by gluon-gluon
components in the $\eta$ and $\eta^{\prime}$ mesons are discussed.
\end{abstract}

\maketitle

\section{Introduction}

In the last years, experiments carried out in B Factories have provided a
large amount of data for a great variety of processes \cite{Bevan:2014iga}.
Among them, measurements of exclusive meson production, in particular,
$e^{+} e^{-}\rightarrow M\,e^{+}e^{-}$ and $e^{+}e^{-}\rightarrow M\,\gamma$
reactions, where $M=\pi,\eta,\eta^{\prime}$, have provided information about
the photon-to-pseudoscalar meson transition form factors (TFFs),
$F_{M\gamma}\left( Q^{2}\right)$, in the spacelike and timelike momentum
transfer regions, respectively. High virtuality data for the
\mbox{pion-$\gamma$} TFF have been obtained by both BABAR~\cite{:2009mc} and
BELLE~\cite{Uehara:2012ag} Collaborations, while BABAR has also measured the
eta- and eta prime-$\gamma$ TFFs~\cite{:2011hk}. These data have to be added
to those previously reported by the CLEO\
Collaboration~\cite{Gronberg:1997fj} for $\pi$-$\gamma,$ $\eta$-$\gamma$ and
$\eta^{\prime}$-$\gamma$ TFFs, as well as those obtained by the L3
Collaboration \cite{Acciarri:1997yx} for the $\eta^{\prime}$-$\gamma$ TFF
and by the CELLO Collaboration \cite{Behrend:1990sr} for the $\pi$-$\gamma$
TFF.

The new experimental results have led to an intense theoretical work. In
fact, from QCD it is seen that the $M$-$\gamma$ TFFs can be computed in
terms of quark and gluon distribution amplitudes (DAs). Moreover, one can
determine the corresponding asymptotic $Q^2\to \infty$ limits, which turn
out to be model independent quantities~\cite{Efremov:1979qk, Lepage:1980fj}.
In the case of the $\pi$-$\gamma$ TFF, the new results
---especially those from the BABAR Collaboration--- indicate that
$F_{\pi\gamma}\left( Q^{2}\right)$ grows with $Q^2$, presumably crossing the
asymptotic QCD limit. The implications of this exciting result have been
widely discussed in the last few years (see \cite{Zhong:2015nxa, Li:2013xna,
deMelo:2013zza, Dumm:2013zoa, Wu:2012kw, Noguera:2012aw, Stefanis:2011kp,
Stefanis:2012yw, deTeramond:2012rt} and references therein). On the other
hand, being less controversial, $\eta$-$\gamma$ and $\eta^{\prime}$-$\gamma$
TFFs have received less theoretical attention. Phenomenological studies have
been carried out in Refs.~\cite{Agaev:2003kb, Kroll:2002nt, Kroll:2013iwa,
Agaev:2014wna}, looking at the gluon content of the $\eta$ and
$\eta^{\prime}$ mesons. Other approaches have been followed in
Refs.~\cite{Escribano:2013kba,Escribano:2015nra}, where the TFFs are
analyzed in a model independent way through the usage of rational Pad{\'e}
approximants, in Ref.~\cite{Klopot:2012hd}, where the anomaly sum rule is
used, and in Ref.~\cite{Czyz:2012nq}, where a formalism based on a chiral
effective theory with two octet resonances is considered. Regarding quark
model approaches, calculations have been carried out within the light front
quark model~\cite{Geng:2012qg} and, for the $\eta$-TFF, within the
Nambu$-$Jona-Lasinio (NJL) model~\cite{Noguera:2011fv}.

In a recent paper \cite{Dumm:2013zoa} we have studied the $\pi$-$\gamma$ TFF
in the framework of a two-flavor version of a nonlocal NJL (nlNJL) quark
model. We extend here our analysis to the case of $\eta$-$\gamma$ and
$\eta^{\prime}$-$\gamma$ TFFs considering a SU(3) flavor version of this
nonlocal effective approach~\cite{Scarpettini:2003fj, Carlomagno:2013ona},
which represents an improvement over the local NJL model. In fact,
nonlocality arises naturally in quantum field theory when the interactions
involve large coupling constants. It can be seen that nonlocal form factors
regularize the model in such a way that anomalies are preserved and charges
are properly quantized, and there is no need to introduce extra cutoffs.
Moreover, our formalism ensures the preservation of fundamental symmetries
(chiral, Poincar\'{e} and local electromagnetic gauge invariances), which
guarantee the proper normalization of the quark DAs.

The quark propagator is taken as one of the main ingredients of our model,
the reason being that lattice QCD (LQCD) calculations allow to obtain
information on this quantity directly from the fundamental QCD theory. These
calculations lead to a definite momentum dependence for both the quark mass
and the quark wave function renormalization~\cite{Parappilly:2005ei,
Bowman2003}. Our model represents, in fact, the minimal framework that
allows to incorporate the corresponding full momentum dependence by choosing
adequate nonlocal form factors~\cite{Noguera:2005ej, Noguera:2005cc,
Noguera:2008}. On the other hand, as it is usual in quark models, our nlNJL
model neglects the explicit presence of gluons when describing the mesonic
states, which are driven by their quark content. Thus, the $\eta$ and
$\eta^{\prime}$ states involve a $q\bar{q}$ octet state (as in the case of
the $\pi$ meson) and a $q\bar{q}$ flavor singlet state. However, one can
also build up a singlet state from two gluons, and the $q\bar{q}$ flavor
singlet components in $\eta$ and $\eta^{\prime}$ mesons will actually become
mixed with the $gg$ component by the $Q^2$ evolution, inducing a two-gluon
contribution at the leading twist order. Consequently, whereas the $\pi$
meson state is described in the TFF calculation by a single DA, for the
$\eta$-$\eta^{\prime}$ sector one needs in general three different DAs, two
of them corresponding to the quark component and one to the gluon component.

One of our objectives will be to analyze the effect of this gluon component.
If we remain faithful to the philosophy of quark models, the latter has to
be neglected. In that case octet and singlet states evolve in a similar way,
and we can perform the $Q^2$ evolution at next-to-leading order (NLO) to
obtain the virtuality dependence of the TFFs. The quark DAs provide the
dominant twist two contribution to the $M$-$\gamma$ TFFs, and corrections to
this leading order can be introduced by considering contributions that carry
extra powers of $1/Q^{2}$ (we include here $1/Q^{4}$ and $1/Q^{6}$ terms).
Therefore, in this scheme we will fix the quark DAs as well as two free
parameters (coefficients of the subleading terms) in the $M$-$\gamma$ TFFs.
Alternatively, we can assume that gluons are present already at low
virtuality, which represents an additional ingredient to our model. In this
second approach we will fit the lowest Gegenbauer coefficients of the gluon
DA using the experimental data.

The present paper is organized as follows. In Sec.~II we develop our
formalism: (A) we describe the connection between $M$-$\gamma$ TFFs and
quark DAs, (B) we present the model Lagrangian and quote our analytical
results for the quark DAs, and (C) we discuss the virtuality dependence of
the DAs through the evolution equations. In Sec.~III we show and discuss the
numerical results for the quark DAs obtained within our model for $\pi$,
$\eta$ and $\eta^{\prime}$ mesons. In Sec.~IV.A we analyze the results
obtained for the $\eta$-$\gamma$ and $\eta^{\prime}$-$\gamma$ TFFs
neglecting the presence of gluons. We also show that if we assume that no
gluons are present at low virtuality, the evolution equations do not
generate a significant presence of gluons at higher $Q^{2}$ values. Then in
Sec.~IV.B we analyze the effect of the presence of gluons at low virtuality
on the description of the $\eta$-$\gamma$ and $\eta^{\prime}$-$\gamma$ TFFs.
Finally, in Sec.~V we present our conclusions. Details of the calculations,
including some relevant analytical expressions, can be found in Appendixes A
and B.

\section{Formalism}

\subsection{Theoretical framework}

The transition form factors for the processes
$M\rightarrow\gamma\gamma^{\ast}$, $M = \eta,\,\eta'$, at large virtuality
$Q^{2}$ are basically determined by the quark and gluon distribution
amplitudes $\Phimq$ and $\phimg$. At the leading order in powers of
$1/Q^{2}$ one has
\begin{align}
Q^{2}\,\Fmg(Q^{2}) \  = \ & \int
dx~\frac{1}{2}\,T_{q\bar{q}}(x,Q^{2},\mu^{2})  ~
\Phimq (x,\mu^{2})
\nonumber\\
&  +\int dx~\frac{1}{2}\,T_{gg}(x,Q^{2},\mu^{2})
\,\frac{4} {3\sqrt{3}}\, f_M^{0}\,\phimg (x,\mu^{2})\ ,
\label{TFF.01}
\end{align}
where $f_M^0$ is a weak decay constant and $T_{q\bar{q}}$, $T_{gg}$ are the
amplitudes of the parton level subprocesses $q\bar{q}\rightarrow\gamma
\gamma^{\ast}$, $gg\rightarrow\gamma\gamma^{\ast}$ evaluated at
next-to-leading order in perturbative QCD. These are given
by~\cite{delAguila:1981nk,Braaten:1982yp,Kroll:2002nt}
\begin{eqnarray}
T_{q\bar{q}}\left(  x,Q^{2},\mu^{2}\right) & = & \frac{1}{x}\left\{
1+C_{F}\frac{\alpha_{s}\left(  \mu\right)  }{4\pi}\left[  \ln^{2}%
x-\frac{x\,\ln x}{1-x}-9+\left(  3+2\ln x\right)  \ln\frac{Q^{2}}{\mu^{2}%
}\right]  \right\} %\nonumber\\
+\left(  x\rightarrow1-x\right) \ , %\label{TFF.04}%
\nonumber \\
T_{gg}\left(  x,Q^{2},\mu^{2}\right)  &  = & C_{F}\frac{\alpha_{s}\left(
\mu\right)  }{4\pi}\,\frac{2\,\ln x}{\left(  1-x\right)  ^{2}}\left[
3-\frac{1}{x}-\frac{1}{2}\ln x-\,\ln\frac{Q^{2}}{\mu^{2}}\right]
%\nonumber\\ &
-\left(  x\rightarrow1-x\right) \ ,
\label{TFF.06}%
\end{eqnarray}
where $C_{F}=4/3$ is a color group factor. As usual, we will
choose the scale $\mu^{2}=Q^{2}$, removing $\ln (Q^2/\mu^2)$
terms.

The function $\Phimq(x,\mu^{2})$ in Eq.~(\ref{TFF.01}) is given by a
combination of quark DAs, which carry the soft, nonperturbative
contributions to the form factor. When studying the evolution of both quark
and gluon DAs it is convenient to write the operators using the SU(3)$_F$
Gell-Mann matrices $\lambda^i$, $i=1,\dots 8$, plus
$\lambda^{0}=\sqrt{2/3}\, I$, while to calculate the quark DAs within quark
models it is usually preferable to choose a flavor basis. Thus we define the
matrix $\lambda^{\ell}=(\sqrt{2}\lambda^{0}+\lambda^{8})/\sqrt{3} = {\rm
diag}(1,1,0)$, which is the identity in the $(u,d)$ flavor subspace, and
$\lambda^{s}={\rm diag}(0,0,\sqrt{2}) =
(\lambda^{0}-\sqrt{2}\lambda^{8})/\sqrt{3}$. In these two basis
$\Phimq(x,\mu^{2})$ is written as
\begin{eqnarray}
\Phimq \left(  x,\mu^{2}\right)   &  = & \frac{4}{3\sqrt{3}} \fmq0  ~\phimqq0 \left(  x,\mu^{2}\right)  +
                                              \frac{\sqrt{2}}{3\sqrt{3}} \fm8 ~ \phimq8 \left(  x,\mu^{2}\right)
\nonumber\\
 &  = & \frac{5\sqrt{2}}{9} \fml ~\phimql   \left(  x,\mu^{2}\right)  +
                                              \frac{2}{9}        \fms ~\phimqs   \left(  x,\mu^{2}\right)\ ,
\label{DAq.01}%
\end{eqnarray}
where the quark DAs are given by
\begin{equation}
\phimqi \left(  x\right)     =-\,\frac{i}{\sqrt{2}\,\fmi}
 \int\frac{dz^{-}}{2\pi}\,e^{iP^{+}z^{-}\left(  x-\frac{1}{2}\right)
}\,\left.  \left\langle 0\right\vert \,\bar{\psi}\left(  -\tfrac{z}{2}\right)
\,\gamma^{+}\gamma_{5}\,\lambda^{i}\,\psi\left(  \tfrac{z}{2}\right)
\,\left\vert M \right\rangle \right\vert _{z^{+}=\vec{z}_{T}=0} \ ,
\label{DAq.02}
\end{equation}
with $i=0,8$ or $i=\ell,s$, depending on the basis choice. We use here
light-front spacetime coordinates $x^\pm = (x^0\pm x^3)/\sqrt{2}$, $\vec x_T
= (x^1,x^2)$. The meson weak decay constants $\fmi$ are defined by
\begin{equation}
\fmi =\frac{1}{i\sqrt{2}\,P^{+}}\,\left\langle 0\right\vert \,\bar
{\psi}\left(  0\right)  \,\gamma^{+}\gamma_{5}\,\lambda^{i}\,\psi\left(
0\right)  \,\left\vert M \right\rangle \ ,
\label{fmes}
\end{equation}
thus it is easy to see that the quark DAs satisfy the sum rule
\begin{equation}
\int_{0}^{1}dx~\phimqi\left(  x\right)  =1
\label{DAq.05}%
\end{equation}
for any scale $\mu$. Moreover, the quark DAs are symmetric under the change
$x\rightarrow\left(  1-x\right)$. Finally, the gluon DA in
Eq.~(\ref{TFF.01}) is given by
\begin{equation}
\phimg\left(  x\right)   =\frac{2}{\sqrt{3} \, \fmq0}
\frac{1}{P^{+}}\int\frac{dz^{-}}{2\pi}\,e^{iP^{+}z^{-}\left(  x-\frac{1}%
{2}\right)  }\,n_{\mu}\,n_{\nu}\,\left.  \left\langle 0\right\vert
G^{\mu\alpha}\left(  -\tfrac{z}{2}\right)  \,\tilde{G}_{\alpha}^{\nu}\left(
\tfrac{z}{2}\right)  \,\left\vert P\right\rangle \right\vert _{z^{+}=\vec
{z}_{T}=0}\ ,
\label{DAg.01}%
\end{equation}
where $G^{\mu\nu}$ is the gluon field strength tensor and
$\tilde{G}^{\mu\nu}=\frac{1}{2}\,\epsilon^{\mu\nu\alpha\beta}
\,G_{\alpha\beta}$. Notice that $\phimg(x)$ is antisymmetric under
the change $x\rightarrow\left(  1-x\right)$, hence
\[
\int_{0}^{1}dx~\phimg\left(  x\right)  =0\ ,
\]
and there is no natural way to normalize the gluon DA. The
prefactor present in Eq.~(\ref{DAg.01}) is just a convention, and
other definitions can be found in literature (see the discussion
in Ref.~\cite{Kroll:2002nt}). A change in this prefactor can be
compensated through a factor into the integrand in the second term
of Eq.~(\ref{TFF.01}).

In the case of the $\pi\rightarrow\gamma\gamma^{\ast}$ TFF the
situation is simpler, since there is no singlet contribution. One
has
\begin{equation}
Q^{2}~F_{\pi\gamma}(  Q^{2}) \ =\ \frac{\sqrt{2}}{3}\,f_{\pi}\int
dx~\frac{1}{2}\,T_{q\bar{q}}(  x) \,\phi_{\pi}(  x,\mu^{2})\ ,
\label{TFF.08}%
\end{equation}
where the pion DA is given by
\begin{equation}
\phi_{\pi}(x) \ =\ \frac{-i}{\sqrt{2}\;f_{\pi}}
\int\frac{dz^{-}}{2\pi}\,e^{iP^{+}z^{-}\left(  x-\frac{1}{2}\right)
}\,\left.  \left\langle 0\right\vert \,\bar{\psi}(-\tfrac{z}{2})
\,\gamma^{+}\gamma_{5}\,\lambda^{3}\,\psi(\tfrac{z}{2})
\left\vert \pi\right\rangle \right\vert _{z^{+}=\vec{z}_{T}=0} \
.
\label{DApi.01}
\end{equation}

\subsection{Neutral pseudoscalar meson distribution amplitudes in a nonlocal NJL model}

We consider here the meson DAs within a nonlocal NJL (nlNJL) model. The
corresponding Euclidean effective action, in the case of SU(3)$_F$ flavor
symmetry, is given by~\cite{Carlomagno:2013ona}
\begin{eqnarray}
\label{se} S_E &=& \int d^4x \ \left\{ \bar \psi(x)(-i\slashed
\partial + \hat m)\psi(x)-\frac{G}{2}\left[
j_a^S(x)j_a^S(x)+j_a^P(x)j_a^P(x)+j^r(x)j^r(x)\right] \right.
\nonumber\\
    &&\qquad \qquad \qquad  \qquad \left. -\frac{H}{4} A_{abc}\left[
j_a^S(x)j_b^S(x)j_c^S(x)-3j_a^S(x)j_b^P(x)j_c^P(x)\right] \right\}
\ ,
\end{eqnarray}
where $\psi(x)$ is the SU(3)$_F$ fermion triplet $\psi = (u\ d\
s)^T$, and $\hat m={\rm diag}(m_u,m_d,m_s)$ is the current quark
mass matrix. We will work in the isospin symmetry limit, assuming
$m_u=m_d$. The model includes flavor mixing through the 't
Hooft-like term driven by $H$, where the constants $A_{abc}$ are
defined by
\begin{equation}
\label{aabc}
A_{abc}\ =\ \frac{1}{3!}\,\epsilon_{ijk}\epsilon_{mnl}(\lambda^a)_{im}
(\lambda^b)_{jn}(\lambda^c)_{kl} \ ,
\end{equation}
with $a=0,\dots ,8$. The fermion currents are given by
\begin{eqnarray}
j_a^s(x) &=& \int d^4z\; \mathcal{G}(z)\,
\bar \psi\left(x+\frac{z}{2}\right)\lambda^a
\psi\left(x-\frac{z}{2}\right)
\ , \nonumber\\
j_a^p(x) &=& \int d^4z\; \mathcal{G}(z)\,
\bar\psi\left(x+\frac{z}{2}\right)\imath \lambda^a \gamma_5
\psi\left(x-\frac{z}{2}\right)
\ , \nonumber\\
j^r(x)   &=& \int d^4z\; \mathcal{F}(z)\,
\bar\psi\left(x+\frac{z}{2}\right) \frac{\imath
\overleftrightarrow{\slashed \partial}}{2\kappa}
\psi\left(x-\frac{z}{2}\right)\ , \label{curr}
\end{eqnarray}
where the functions $\mathcal{G}(z)$ and $\mathcal{F}(z)$ are covariant form
factors responsible for the nonlocal character of the interactions. Notice
that the relative weight of the interaction driven by $j^r(x)$, which leads
to quark wave function renormalization, is controlled by the parameter
$\kappa$. In the mean field approximation (MFA), which will be used here in what
follows, the quark propagator for each flavor $f=u,d,s$ can be expressed as
\begin{eqnarray}
S_f(p) = \frac{Z(p)}{-\slashed p + M_f(p)}\ , \label{quarkp}
\end{eqnarray}
where $M_f(p)$ and $Z(p)$ stand for momentum dependent effective mass and
wave function renormalization (WFR), respectively. One
has~\cite{Carlomagno:2013ona}
\begin{equation}
M_f(p) \ =\  Z(p)\, \bigg[ m_f\, +\, \bar \sigma_f\, g(p)\bigg] \ ,
\qquad \  Z(p) \ =\ \left[ 1\,-\,\frac{\bar \zeta}{\kappa}\,f(p)\right]^{-1}\
, \label{mz}
\end{equation}
where $g(p)$ and $f(p)$ are Fourier transforms of $\mathcal{G}(z)$ and
$\mathcal{F}(z)$, while $\bar \sigma_f$ and $\bar \zeta$ are mean field
values of scalar fields associated with the corresponding currents in
Eq.~(\ref{curr}). Details of the procedure carried out to obtain these
quantities are given in App.~A.

The momentum dependence of the interaction form factors can be now obtained
from lattice QCD results. Following the analysis in Ref.~\cite{Bowman2003},
the effective mass $M_u(p)$ can be written as
\begin{equation}
M_u(p) \ = \ m_u \, + \, \alpha_m\, f_m(p) \ ,
\end{equation}
where
\begin{equation}
f_m(p) \ = \ 1/\left[ 1 + (p^2/\Lambda_0^2)^\alpha \right]\ ,
\label{fm}
\end{equation}
with $\alpha = 3/2$. From Eqs.~(\ref{mz}) one has then $\alpha_m =
(m_u\bar\zeta/\kappa +\bar\sigma_u)/(1-\bar\zeta/\kappa)$. For the
wave function renormalization we use the
parametrization~\cite{Noguera:2005ej,Noguera:2008}
\begin{equation}
Z(p) \ = \ 1 \, - \, \alpha_z\, f_z(p) \ ,
\end{equation}
where
\begin{equation}
f_z(p) \ = \ 1/\left(1 + p^2/\Lambda_1^2\right)^{5/2}\ .
\label{fz}
\end{equation}
Here the new parameter $\alpha_z$ is given by $\alpha_z =
-\bar\zeta/(\kappa-\bar\zeta)$. The functions $f(p)$ and $g(p)$ can be now
easily obtained from Eqs.~(\ref{mz}-\ref{fz}). As shown in
Refs.~\cite{Noguera:2005ej,Noguera:2008}, for an adequate choice of
parameters these functional forms can reproduce very well the momentum
dependence of quark mass and WFR obtained in lattice calculations. We
complete the model parameter fixing by taking as phenomenological inputs the
values the of the pion, kaon and $\eta'$ masses and the pion weak decay
constant~\cite{Carlomagno:2013ona}. The resulting model parameters are given
in Table~\ref{Table.param}.

\begin{table} [h]
\begin{center}
\begin{tabular}{ccccccc}
\hline \hline \hspace{0.5cm} $m_u$ (MeV) \hspace{0.5cm} & $m_s$ (MeV) &
\hspace{0.5cm}$G\Lambda_0^2$\hspace{0.5cm} &
\hspace{0.5cm}$-H\Lambda_0^5$\hspace{0.5cm} &
\hspace{0.5cm}$\kappa/\Lambda_0$\hspace{0.5cm} & $\Lambda_0$ (GeV) & \hspace{0.5cm} $\Lambda_1$ (GeV) \hspace{0.5cm} \\
 \hline
 2.6  & 64.9  &     16.65      &     202.8       &  10.34   &   0.795     &    1.510      \\
\hline \hline
\end{tabular}
\end{center}
\caption{Model parameters}
\label{Table.param}
\end{table}

Given this effective model for the strong interactions at low energies, one
can explicitly evaluate the quark DAs from Eq.~(\ref{DAq.02}). Since the
amplitude involves a bilocal axial vector current, one should introduce into
the effective action in Eq.~(\ref{se}) a coupling to an external axial gauge
field $\mathcal{A}^a_{\mu}$. For a local theory this can be done just
through the replacement
\begin{equation}
\partial_{\mu}\ \rightarrow\ \partial_{\mu}+i\ \gamma_{5}\, \lambda^a \
\mathcal{A}^a_{\mu}(y)\ .
\end{equation}
In the case of the above described nonlocal model, however, the situation is
more complicated since the inclusion of gauge interactions implies a change
not only in the kinetic piece of the Lagrangian but also in the nonlocal
currents appearing in the interaction terms. If $x$ and $z$ denote the space
variables in the definitions of the nonlocal currents [see
Eq.~(\ref{curr})], one has
\begin{align}
\psi(x-z/2)\  &  \rightarrow W\left(  x,x-z/2\right)  \ \psi
(x-z/2)\ ,\nonumber\\
\psi^{\dagger}(x+z/2)\  &  \rightarrow\psi^{\dagger}(x+z/2)\
W\left(
x+z/2,x\right)  \ . \label{gauge}
\end{align}
Here the function $W(s,t)$ is defined by
\begin{equation}
W(s,t)\ =\ \mathrm{P}\;\exp\left[  i\ \int_{s}^{t}dr_{\mu}\
\gamma_{5}\, \lambda^a \ \mathcal{A}^a_{\mu}(r)\right]  \ ,
\label{intpath}
\end{equation}
where $r$ runs over an arbitrary path connecting $s$ with $t$.

This procedure has been analyzed in detail within nlNJL models, in
particular regarding the calculation of the pseudoscalar meson decay
constants~\cite{Scarpettini:2003fj,Noguera:2005ej,Bowler:1994ir},
see Eq.~(\ref{fmes}). The situation is similar for the case
of the bilocal axial current in the definition of the meson DA. In
fact, the basic physical idea beyond the factorization of the
meson TFF into hard and soft contributions is that for high $Q^{2}$
the struck quark loses its high momentum before being able to
interact with the remaining quarks and gluons of the hadron
($Q^{2}\sim1$ GeV$^{2}$ implies a time scale of the order of
$10^{-24}$ s). Therefore, the nonlocal interaction does not see
the struck quark but only the quarks in the hadron before and
after the photon absorption-emission process. This can be
effectively implemented by introducing an external fictitious
probe carrying the adequate quantum numbers, which in our case is
a gauge axial field (a similar situation has been studied in the
case of the pion parton distribution, see
Refs.~\cite{Noguera:2005ej,Noguera:2005cc}). Thus, the axial
vertex in Eq.~(\ref{DAq.02}) will become dressed by the
nonlocal interaction, irrespective of whether the quark current is
a local or a bilocal one (as in this case).

The steps to be followed in the explicit calculation of the
quark DA within the nlNJL model are detailed in
Appendix~A. We quote here the resulting expression.
In the flavor basis (i.e. $q_i=q_\ell,q_s$) we have
\begin{equation}
\phimqi(x)  =\frac{2\sqrt{2}\,N_{c}\,g_{M qq}}{\fmi}%
\int\frac{dw\,d^{2}k_{T}}{\left(  2\pi\right)  ^{4}}\
F_i\left(  w,x,k_{T} \right)  \ , \label{DA.01}
\end{equation}
where $g_{M qq}$ stands for an effective quark-meson coupling constant [see
Eq.~(\ref{gmqq}) in Appendix~A] and the integration variables are related to
the meson and quark Euclidean four-momentum $P$ and $k$, respectively.
Considering the light front variables in the frame where the transverse
component ${\vec{P}}_{T}$ vanishes, the invariants $k^{2}$ and $k\cdot P$
can be written in terms of the variables $w$ and $k_{T}$ as
\[
k^{2}=-i\,w\left(  x-\frac{1}{2}\right)  +m_{M}^{2}\left(  x-\frac{1}%
{2}\right)  ^{2}+k_{T}^{2}\ ,\qquad k\cdot P=-i\,\frac{w}{2}\ .
\]
It is convenient to separate the integrand in Eq.~(\ref{DA.01})
into two pieces,
\begin{equation}
F_i\left(  w,x,k_{T}\right)  \ =
\ F_i^{(1)} \left(  w,x,k_{T}\right) + F_i^{(2)}\left( w,x,k_{T}\right)  \ .
\label{terms}
\end{equation}
The explicit expressions for these functions are
\begin{eqnarray}
F_i^{(1)}\left(  w,x,k_{T}\right)  &  = & \frac{g(k)}{2}
\frac{Z(k_{+})\,Z(k_{-})}{D_i(k_{+})D_i(k_{-})}
\left[
\frac{1}{Z(k_{+})}+\frac{1}{Z(k_{-})}\right]
\left[  \left( 1-x\right)  \, M_i(k_{+})+ x \, M_i(k_{-})\right]  \ ,
\label{DA.10}\\
F_i^{(2)}\left(  w,x,k_{T}\right) \rule{0cm}{0.9cm}  &  = &
g(k)\frac{Z(k_{+})Z(k_{-})}{D_i(k_{+})D_i(k_{-})}
\Big\{[k_{+}\cdot k_{-}+ M_i(k_{+})\, M_i(k_{-})]\,\nu_i^{(1)} \nonumber\\
& & -\; k\cdot\left[  k_{+} \,  M_i(k_{-}) - k_{-} M_i(k_{+})\right] \, \nu^{(2)}\Big\}
-\frac{M_i(k) \, Z(k)}{D_i(k)\,\bar{\sigma}_{i}}\ \nu_i^{(1)}\
,
\label{DA.11}%
\end{eqnarray}
where $M_\ell=M_u=M_d$ and $\bar \sigma_\ell = \bar \sigma_u = \bar
\sigma_d$. We have defined $k_{\pm}=k\pm P/2$ and $D_i(k)=k^{2}+
M_i(k)^{2}$, while the functions $\nu^{(1)}_i$ and $\nu^{(2)}$ in $F_i^{(2)}$
are given by
\begin{align}
\nu_i^{(1)}  &  =\frac{\left(  x-\frac{1}{2}\right)  }{k\cdot P}
\Big[  \frac{M_i(k_{+})}{Z(k_{+})}  +  \frac{M_i(k_{-})}{Z(k_{-})} -
2 \frac{M_i(k)}{Z(k)} + m_M^{2}\,\bar{\sigma}_{i}\,\alpha_{g}^{-}\Big]+ \bar{\sigma}_{i}%
\,\alpha_{g}^{-}\ ,\nonumber\\
\nu^{(2)}  &  =\frac{\left(  x-\frac{1}{2}\right)  }{k\cdot P}
\Big[\frac{1}{Z(k_{-})}-\frac{1}{Z(k_{+})} + m_M^{2}\, {\bar \zeta}
\,\alpha_{f}^{+}\Big]+{\bar \zeta}\,\alpha_{f}^{+}\ . \label{nu12}%
\end{align}
Here $\alpha_{g}^{-}$ and $\alpha_{f}^{+}$ depend in general on the
integration path in Eq.~(\ref{intpath}). For a straight line path one has
\begin{eqnarray}
\alpha_{g}^{-} & = &
\int_{0}^{1}d\lambda\,\frac{\lambda}{2} \ g^{\,\prime}(k-\lambda P/2)-
\int_{-1}^{0}d\lambda\ \frac{\lambda}{2}\
g^{\,\prime}(k-\lambda P/2)\ , \nonumber\\
\alpha_{f}^{+} & = & \int_{-1}^{1}\
d\lambda\ \frac{\lambda }{2}\ f^{\,\prime}(k-\lambda P/2)\ ,
\label{alphas}
\end{eqnarray}
where $g'(k) \equiv dg(k)/dk^2$, and same for $f'(k)$.

It is important to mention that, even when our effective model leads to an
adequate phenomenological pattern for low energy meson phenomenology, there
are some differences between model predictions and phenomenological values
of the $\eta$ and $\eta'$ decay constants (see Table~\ref{Table.pheno} in
App.~A). In our numerical calculations, when evaluating the $\eta$ and
$\eta'$ DAs from Eq.~(\ref{DA.01}) we will take the values of $f_M^i$
arising from our model, in order to guarantee the proper normalization
condition Eq.~(\ref{DAq.05}). On the other hand, we will use the
phenomenological values for $f_M^\ell$ or $f_M^i$ when evaluating the flavor
mixing leading to the quark DAs, Eq.~(\ref{DAq.01}).

\subsection{Distribution amplitude evolution}
\label{evolution}

Let us analyze the evolution of the DAs with the energy scale. Firstly,
notice that QCD evolution equations mix the $q\bar{q}$ singlet flavor
component with the $gg$ component in $\eta$ and $\eta^{\prime}$ DAs.
Consequently, after obtaining the low energy $q\bar{q}$ flavor DAs
$\phimqi$, $i=\ell,s$, from the effective quark model, it is convenient to
change to the octet and singlet DAs
\begin{align}
\phimq8 (x) \  &
=\ \frac{1}{\sqrt{3}~\fm8  }
 \left[ \fml~\phimql(x)  -\sqrt{2}~\fms~\phimqs(x)
 \right]\ ,
\nonumber\\
\phimqq0 (x)  \ &
=\ \frac{1}{\sqrt{3} ~\fmq0}
 \left[  \sqrt{2}~\fml~\phimql(x) + \fms~ \phimqs(x)
 \right]\ .
\label{DAEv.01}
\end{align}
Once the latter are known at a given $\mu_{0}$ scale, their evolution up to
a higher scale $\mu$ can be obtained from perturbative QCD. In order to
study this evolution it is convenient to expand the DAs in series of
Gegenbauer polynomials:
\begin{align}
\phimqi(  x,\mu)  \ &  =\ 6\,x\,(  1-x)  \,\sum
_{n=0,2,4,...} \aqimn \left(  \mu\right)  ~C_{n}^{3/2}\left(  2x-1\right) \
,
\nonumber\\
\phimg(  x,\mu) \  &  =\ x^{2}\,\left(  1-x\right)  ^{2}%
\,\sum_{n=2,4,...}\agmn\left(  \mu\right)  ~C_{n-1}^{5/2}\left(
2x-1\right)\ ,
\label{DAEv.03}
\end{align}
where $i=0,8$, and we have now explicitly denoted the $\mu$ dependence of
the DAs. Notice that only $n$-even terms contribute to the sums, due to the
symmetric (antisymmetric) behavior of the quark DAs (gluon DA) under the
replacement $x\leftrightarrow 1-x$. Moreover, since $\phimqi(x,\mu)$
($i=0,8$) satisfy the sum rule Eq.~(\ref{DAq.05}), the first coefficients
$a^{(q_0)}_{M0}\left( \mu\right)$ and $a^{(q_8)}_{M0}\left(\mu\right)$ have
to be equal to 1 for any value of $\mu$. Thus, all the information from the
meson effective model is included in the remaining coefficients
$\aqimn(\mu)$ and $\agmn(\mu)$, with $n=2,4,\dots$.

From the orthogonality relations satisfied by the Gegenbauer polynomials one
can obtain the coefficients at the $\mu_{0}$\ scale, namely
\begin{align}
\aqimn\left(  \mu_{0}\right)   &  =\frac{2\left(  2n+3\right)  }{3\left(
n+1\right)  \left(  n+2\right)  }\int_{0}^{1}dx~C_{n}^{3/2}\left(
2x-1\right)  ~\phimqi\left(  x,\mu_{0}\right) \ ,\label{DAEv.06}\\
\agmn\left(  \mu_{0}\right)   &  =\frac{144\left(  2n+5\right)  }{\left(
n+1\right)  \left(  n+2\right)  (n+3)(n+4)}\int_{0}^{1}dx~C_{n}^{5/2}\left(
2x-1\right)  ~\phimg\left(  x,\mu_{0}\right) \ .\label{DAEv.06.2}
\end{align}
Notice that Eq.~(\ref{DAEv.06}) holds either if one is working in the flavor
basis ($i=\ell,s$) or in the SU(3)$_{F}$ basis ($i=0,8$). At the LO the
Gegenbauer polynomials are eingenfunctions of the evolution kernel,
therefore $a_{Mn}$ coefficients of different order $n$ do not mix with each
other~\cite{Kroll:2013iwa}. On the other hand, as stated, QCD evolution
equations mix the gluon and singlet quark components for $n\geq2$. The
evolution of these coefficients up to a scale $\mu$ is given by (see
Refs.~\cite{Kroll:2002nt,Kroll:2013iwa})
\begin{align}
\aqqq0mn\left(  \mu\right)   &  =\aplmn\left(  \mu
_{0}\right)  \left(  \frac{\alpha_{s}\left(  \mu_{0}\right)  }{\alpha
_{s}\left(  \mu\right)  }\right)  ^{\gamma_{n}^{\left(  +\right)  }/\beta_{0}%
}+\rho_{n}^{\left(  -\right)  }~\amimn\left(  \mu
_{0}\right)  \left(  \frac{\alpha_{s}\left(  \mu_{0}\right)  }{\alpha
_{s}\left(  \mu\right)  }\right)  ^{\gamma_{n}^{\left(  -\right)  }/\beta_{0}%
}~,\nonumber\\
\agmn\left(  \mu\right)   &  =\rho_{n}^{\left(  +\right)  }%
~\aplmn\left(  \mu_{0}\right)  \left(  \frac{\alpha
_{s}\left(  \mu_{0}\right)  }{\alpha_{s}\left(  \mu\right)  }\right)
^{\gamma_{n}^{\left(  +\right)  }/\beta_{0}}+\amimn
\left(  \mu_{0}\right)  \left(  \frac{\alpha_{s}\left(  \mu_{0}\right)
}{\alpha_{s}\left(  \mu\right)  }\right)  ^{\gamma_{n}^{\left(  -\right)
}/\beta_{0}}~.
\label{DAEv.08}
\end{align}
Here $\beta_{0}=11-2\,n_{f}/3$, $n_{f}$ being the number of active
flavors at the scale of the process (in our case we take $n_f = 4$),
and $\gamma_{n}^{(\pm)}$ are the
eigenvalues of the anomalous dimension matrix $\gamma_n$. These are given by
\begin{equation}
\gamma_{n}^{\left(  \pm\right)  }=\frac{1}{2}\left[  \gamma_{n}^{qq}%
+\gamma_{n}^{gg}\pm\sqrt{\left(  \gamma_{n}^{qq}-\gamma_{n}^{gg}\right)
^{2}+4~\gamma_{n}^{qg}~\gamma_{n}^{gq}}\right]  ,
\end{equation}
where the (LO) matrix elements of $\gamma_n$ read % (in our convention)
\begin{align}
\gamma_{n}^{qq}  &  =C_{F}\left[  3+\frac{2}{\left(  n+1\right)  \left(
n+2\right)  }-4\sum_{i=1}^{n+1}\frac{1}{i}\right]  ~,\nonumber\\
\gamma_{n}^{qg}  &  =C_{F}\frac{n~\left(  n+3\right)  }{3\left(  n+1\right)
\left(  n+2\right)  }~,\\
\gamma_{n}^{gq}  &  =\frac{36}{\left(  n+1\right)  \left(  n+2\right)
}~,\nonumber\\
\gamma_{n}^{gg}  &  =\beta_{0}+N_{c}\left[  \frac{8}{\left(  n+1\right)
\left(  n+2\right)  }-4\sum_{i=1}^{n+1}\frac{1}{i}\right] \ .\nonumber
\end{align}
The coefficients $\rho_{n}^{(+)}$ and $\rho_{n}^{(-)}$, which weight the
presence of quarks in the gluon DA and gluons in the singlet quark DA,
respectively, are given by
\begin{align}
\rho_{n}^{\left(  +\right)  }  &  =6\frac{\gamma_{n}^{gq}}{\gamma_{n}^{\left(
+\right)  }-\gamma_{n}^{gg}}\ ,\nonumber\\
\rho_{n}^{\left(  -\right)  }  &  =\frac{1}{6}\frac{\gamma_{n}^{qg}}%
{\gamma_{n}^{\left(  -\right)  }-\gamma_{n}^{qq}} \ .
\end{align}
Finally, the evolution of the strong coupling constant $\alpha_s$ at the
LO is given by
\begin{equation}
\alpha_{s}(\mu) \ = \ \frac{4\pi}{\beta_{0}\ln(\mu^{2}/\Lambda^{2})}\ ,
\end{equation}
with $\Lambda=0.224$~GeV.

In Table~\ref{Table.gammas} we quote the values of the anomalous dimensions
for the first values of $n$. Already for $n=2$ it is seen that
$\gamma^{(+)}$ and $\gamma^{(-)}$ are close to $\gamma^{qq}$ and
$\gamma^{gg}$, respectively, and the differences become even smaller for
larger $n$. The numerical values for $\rho^{(\pm)}$ and the product
$\rho_{n}^{(+)}\rho_{n}^{(-)}$ for the first values of $n$ are given in
Table~\ref{Table.RhoMm}.
\begin{table}[tbp] \centering
\begin{tabular}
[c]{|c|c|c|c|c|c|c|}\hline
$n$ & $2$ & $4$ & $6$ & $8$ & $10$ & Asymptotic form\\\hline
$\gamma_{n}^{\left(  +\right)  }$ & $-5.379$ & $-8.040$ & $-9.759$ & $-11.046$
& $-12.078$ & $-\tfrac{16}{3}\ln n$\\
$\gamma_{n}^{\left(  -\right)  }$ & $-11.84$ & $-18.32$ & $-22.37$ & $-25.36$
& $-27.73$ & $-12\,\ln n$\\
$\gamma_{n}^{qq}$ & $-5.556$ & $-8.089$ & $-9.781$ & $-11.06$ & $-12.09$ &
$-\tfrac{16}{3}\ln n$\\
$\gamma_{n}^{gg}$ & $-11.67$ & $-18.27$ & $-22.35$ & $-25.35$ & $-27.72$ &
$-12\,\ln n$\\\hline
\end{tabular}
\caption{Numerical values of the first $ \gamma ^{( \pm )} _{n}$, $ \gamma
^{ qq} _{n}$ and  $ \gamma ^{gg} _{n}$ coefficients, and asymptotic values.}
\label{Table.gammas}
\end{table}

\begin{table}[tbp] \centering
\begin{tabular}
[c]{|c|c|c|c|c|c|c|}\hline
$n$ & $2$ & $4$ & $6$ & $8$ & $10$ & Asymptotic form\\\hline
$\rho_{n}^{\left(  +\right)  }$ & $2.8627$ & $0.7041$ & $0.3063$ & $0.1678$ &
$0.1045$ & $162/(5\,n^{2}\,\ln n)$\\
$\rho_{n}^{\left(  -\right)  }$ & $-0.0098$ & $-0.0068$ & $-0.0057$ &
$-0.0051$ & $-0.0047$ & $-1/(90\,\ln n)$\\
$\rho_{n}^{\left(  +\right)  }\rho_{n}^{\left(  -\right)  }$ & $-0.0281$ &
$-0.0048$ & $-0.0017$ & $-0.0008$ & $-0.0005$ & $-9/(25\,n^{2}\,\ln^{2}%
n)$\\\hline
\end{tabular}
\caption{Numerical values of the first $ \rho ^{( \pm )} _{n}$ coefficients,
and asymptotic values.} \label{Table.RhoMm}
\end{table}

In the case of the distribution amplitude $\phimq8$, at the LO the evolution
is just governed by the anomalous dimension $\gamma_{n}^{qq}$. One has
\begin{equation}
\aqq8mn\left(  \mu\right)  =\aqq8mn\left(  \mu_{0}\right)  \left(
\frac{\alpha_{s}\left(  \mu_{0}\right)  }{\alpha_{s}\left(  \mu\right)
}\right)  ^{\gamma_{n}^{qq}/\beta_{0}} \ .\label{DAEv.15}%
\end{equation}

We will also take into account the effect of NLO corrections to the DAs. In
general, at the NLO the evolution equations for different coefficients
$\aqimn$ get mixed, and the pattern becomes more complicated. We will
consider the NLO evolution for the octet component (see discussion in the
next section). The corresponding coefficients evolve according
to~\cite{Agaev:2010aq}
\begin{equation}
{\aqq8mn}^\text{NLO}(\mu)\ =\ \aqq8mn(\mu_{0})\,E_{n}^{\text{NLO}}(\mu,\mu
_{0})+\frac{\alpha_{s}(\mu)}{4\pi}\sum_{k=0}^{n-2}\ \aq8mk(\mu_{0})\,\left(
\frac{\alpha_{s}\left(  \mu_{0}\right)  }{\alpha_{s}\left(  \mu\right)
}\right)  ^{\gamma_{k}^{qq}/\beta_{0}}\,d_{n}^{k}(\mu,\mu_{0})\ .
\label{anNLO.01}%
\end{equation}
Explicit expressions for the renormalization factors $E_{n}^{\text{NLO}}
(\mu,\mu_{0})$, as well as for the off-diagonal mixing coefficients
$d_{n}^{k}(\mu,\mu_{0})$ in the $\overline{\text{MS}}$ scheme are collected
in Appendix~B. Usually the calculation of a few coefficients $\aqimn(\mu)$
is sufficient to get a good estimate of the $\pi$DA at the scale $\mu$ from
Eq.~(\ref{DAEv.03}).

\section{Distribution Amplitudes and Transition Form Factors in the nonlocal NJL model}

\subsection{Quark DAs}

\label{SectionDAs}%

\begin{figure}
[ptb]
\begin{center}
\includegraphics[width=1.0\textwidth]{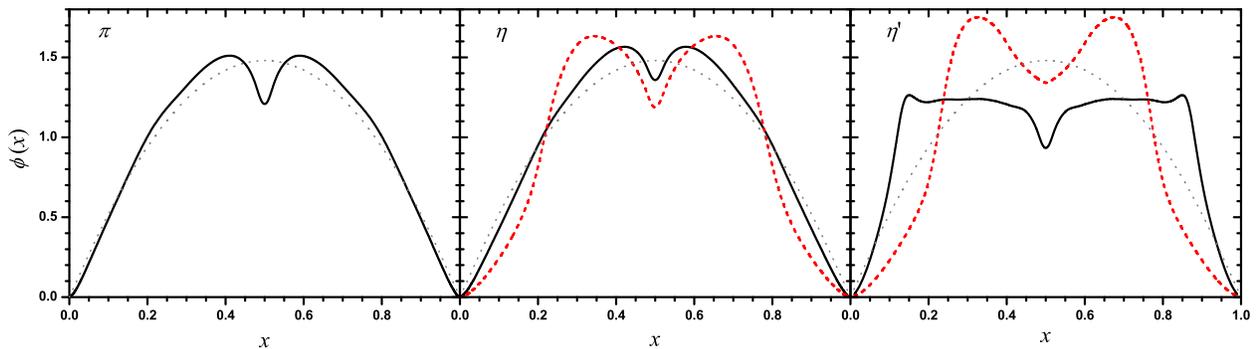}
\caption{Quark distribution amplitudes for the $\pi$, $\eta$ and
$\eta^\prime$ mesons. Left panel: $\pi$ DA, $\phi_{\pi}(x)$. Central panel:
$\eta$ DAs, $\phietl(x)$ (solid line) and $\phiets(x)$ (dashed line). Right
panel: $\eta^\prime$ DAs, $\phietpl(x)$ (solid line) and $\phietps(x)$
(dashed line). The dotted lines correspond in all cases to the asymptotic
limit $\phi_{\rm asym}(x)=6\,x(1-x)$.} \label{FigDAs}
\end{center}
\end{figure}

From the numerical evaluation of the integrals in Eq.~(\ref{DA.01}) one can
obtain the quark DAs for $\pi$, $\eta$ and $\eta'$ mesons within the above
described three-flavor nlNJL model. The corresponding curves are displayed
in Fig.~\ref{FigDAs}, where for comparison we also include the asymptotic
limit $\phi_{\rm asym}(x)=6x(1-x)$. As stated in the previous section, our
calculations have been performed in Euclidean space. The consistency of our
procedure has been discussed in Ref.~\cite{Dumm:2013zoa}, where the pion DA
and TFF are analyzed within a two-flavor version of the model. Our main test
in this sense is the verification that the sum rule Eq.~(\ref{DAq.05}) is
satisfied. In the case of $\eta$ and $\eta'$ mesons, however, the stringency
of this test becomes weakened owing to the numerical uncertainties in the
calculations. In fact, when computing the integrals in Eq.~(\ref{DA.01}) one
has to take into account that the functions $F_i(w,x,k_T)$ show cuts in the
complex $w$ plane that require a deformation of the integration paths (see
e.g.\ the discussion in App.~B of Ref.~\cite{Villafane:2016ukb}). In
addition, depending on the value of $x$ these functions have poles that also
need to be compensated numerically. The normalization of quark DAs obtained
from our calculations in the present model are 1.0004 for $\phi_\pi(x)$,
0.9989 and 0.9753 for $\phietl(x)$ and $\phiets(x)$, respectively, and 1.027
and 0.870 for $\phietpl(x)$ and $\phietps(x)$, respectively. It is worth
mentioning that the effect of poles and cuts gets increased for higher quark
and meson masses, therefore numerical uncertainties are particularly large
in the case of the $s$ quark DA in the $\eta'$ meson, where we find the
largest departure from the normalization condition (the effect of the error
in the determination of the $\eta^{\prime}$ DAs is discussed above). All
quark DAs shown in Fig.~\ref{FigDAs} have been renormalized so that they
satisfy the sum rule.

When using a quark model to describe the deep structure of hadrons it
is crucial to establish the chosen scale $\mu_{0}$ that will be associated
with the results provided by the model. In our case the scale should be the
same as that used in lattice calculations, since we have taken into account
lattice results in order to fix the shape of the form factors in the quark
propagators. Thus, from Ref.~\cite{Parappilly:2005ei} we take $\mu_{0}=3$
GeV, which is a rather large value in comparison with the scale
$\mu_{0}\sim0.5-1$~GeV usually adopted in quark model calculations.

In the left panel of Fig.~\ref{FigDAs} we show the pion DA. Our result is
pretty similar to that obtained within the two-flavor nonlocal NJL model
studied in Ref.~\cite{Dumm:2013zoa}. Notice that this might not have been
the case, since the change from a two-flavor model to the present
three-flavor model implies a refitting of all model parameters. By looking
at the DAs in Fig.~\ref{FigDAs} it is seen that in all cases the curves have
two symmetric maxima. This is also shown by the $\pi$ DAs calculated in
Refs.~\cite{Chernyak:1981zz,Bakulev:2001pa}, albeit in our case the two
maxima are much closer to $x=0.5$. In fact, in the nlNJL model this feature
arises from the term $F_i^{(2)}(w,x,k_T)$ [see
Eqs.~(\ref{terms}-\ref{DA.11})], which is a genuine nonlocal contribution.
Now, by comparing the curves in the different panels of Fig.~\ref{FigDAs}
one can see the effect of meson and quark masses in the behavior of the DAs.
As expected, the $\pi$ DA at $\mu_{0}=3$~GeV is relatively close to the
asymptotic limit $\phi_{\rm asym}(x)=6x(1-x)$. This also holds for the
$u$ (or $d$) quark DA in the case of the $\eta$ meson, $\phietl(x)$, while
for the strange quark DA $\phiets(x)$ the deviation from $\phi_{\rm
asym}(x)$ is more appreciable. Finally, in the case of the $\eta^{\prime}$
meson (right panel in Fig.~\ref{FigDAs}), both $\phietpl\left( x\right)$ and
$\phietps\left( x\right)$ lie far from the asymptotic limit. Another
important feature, common to all obtained DAs, is that they go to zero
rather fast near the points $x=0$ and $x=1,$ supporting the idea of
suppression of the kinematic end points
\cite{Mikhailov:2009kf,Mikhailov:2010ud}.

Let us consider the QCD evolution of the DAs. We recall that we are working
within a quark model in which there is no gluon content. Moreover, according
to the numerical values of the $\rho_{n}^{(\pm)}$ coefficients in
Table~\ref{Table.RhoMm} (which measure the degree of mixing between quark
and gluon components of the DAs in the evolution equations) we can assume
the contribution of gluons to be negligible at any $\mu$ scale. Thus it is
possible to use just the octet evolution formulae for all quark DAs. In
Tables~\ref{Table.an.pi}, \ref{Table.an.eta} and \ref{Table.an.etaprime} we
quote the first coefficients of the Gegenbauer expansion obtained from the
quark DAs at $\mu_{0}=3$ GeV in the flavor basis, together with the
corresponding values after evolving down to $\mu=1$ GeV at NLO. Notice that,
in general, within our approach the absolute values of the expansion
coefficients $\aqimn$ decrease rather slowly with $n$.

\begin{table}[htbp] \centering
\begin{tabular}
[c]{|c|c|c|c|c|c|c|}\hline
$n$ & $2$ & $4$ & $6$ & $8$ & $10$ & $12$\\\hline
\multicolumn{1}{|l|}{$a_{\pi n}(3\,$GeV)} & $-0.0183$ & $-0.0324$ & $0.0048$ &
$-0.0090$ & $0.0049$ & $-0.0067$\\
\multicolumn{1}{|l|}{$a_{\pi n}(1\,$GeV) (NLO)} & $-0.0225$ & $-0.0646$ &
$0.0075$ & $-0.0242$ & $0.0114$ & $-0.0205$\\\hline
\end{tabular}
\caption{Coefficients  $a_{\pi n}$ obtained within the nlNJL model at
$\mu _{0}=3$ GeV, and their values after evolving down to $\mu =1$~GeV at
NLO.} \label{Table.an.pi}
\end{table}
\begin{table}[htbp] \centering
\begin{tabular}
[c]{|c|c|c|c|c|c|c|}\hline
$n$ & $2$ & $4$ & $6$ & $8$ & $10$ & $12$\\\hline
\multicolumn{1}{|l|}{$\aetnql(3\,$GeV)} & $-0.0538$ & $-0.0263$ &
$0.0049$ & $-0.0071$ & $0.0033$ & $-0.0049$\\
\multicolumn{1}{|l|}{$\aetnql(1\,$GeV) (NLO)} & $-0.0778$ & $-0.0540$ &
$0.0077$ & $-0.0194$ & $0.0071$ & $-0.0152$\\\hline
\multicolumn{1}{|l|}{$\aetnqs(3\,$GeV)} & $-0.1185$ & $-0.0577$ &
$0.0538$ & $-0.0248$ & $0.0012$ & $0.0023$\\
\multicolumn{1}{|l|}{$\aetnqs(1\,$GeV) (NLO)} & $-0.1785$ & $-0.1168$ &
$0.1151$ & $-0.0619$ & $0.0016$ & $0.0054$\\\hline
\end{tabular}
\caption{Coefficients  $\aetnql$ and $\aetnqs$ obtained within the nlNJL
model at $\mu _{0}=3$ GeV, and their values after evolving down to $\mu
=1$~GeV at NLO.} \label{Table.an.eta}
\end{table}
\begin{table}[htbp] \centering
\begin{tabular}
[c]{|c|c|c|c|c|c|c|}\hline
$n$ & $2$ & $4$ & $6$ & $8$ & $10$ & $12$\\\hline
\multicolumn{1}{|l|}{$\aetpnql$($3$ GeV)} & $0.1156$ & $-0.0789$
& $-0.0341$ & $0.0023$ & $0.0201$ & $-0.0061$\\
\multicolumn{1}{|l|}{$\aetpnql(1\,$GeV) (NLO)} & $0.1860$ &
$-0.1509$ & $-0.0799$ & $0.0019$ & $0.0520$ & $-0.0182$\\\hline
\multicolumn{1}{|l|}{$\aetpnqs(3\,$GeV)} & $-0.1343$ & $-0.0568$
& $0.0632$ & $-0.0334$ & $0.0010$ & $0.0104$\\
\multicolumn{1}{|l|}{$\aetpnqs(1\,$GeV) (NLO)} & $-0.2031$ &
$-0.1155$ & $0.1360$ & $-0.0829$ & $0.0008$ & $0.0291$\\\hline
\end{tabular}
\caption{Coefficients  $\aetpnql$ and $\aetpnqs$ obtained within the
nlNJL model at $\mu _{0}=3$ GeV, and their values after evolving down to
$\mu =1$~GeV at NLO.} \label{Table.an.etaprime}
\end{table}

Our results for the case of the $\eta$ meson can be compared with those
obtained within the (local) Nambu$-$Jona-Lasinio model in
Ref.~\cite{Noguera:2011fv}, where only the $\eta$ meson case is analyzed,
since the $\eta^{\prime}$ turns out to be unbounded. It is seen that the
shapes of the $\eta$ DAs are quite different from those obtained in the
present model, showing only one central maximum (we recall that the origin
of the two-maxima behavior shown in Fig.~\ref{FigDAs} arises from the
purely nonlocal contribution). As expected, the differences in the shapes
are translated to the coefficients of the Gegenbauer expansion: the first
coefficients obtained in Ref.~\cite{Noguera:2011fv} read
$a_{\eta2}^{(q_\ell)}=0.134,$ $a_{\eta4}^{(q_\ell)}=0.352,$
$a_{\eta2}^{(q_s)}=0.377$ and $a_{\eta4}^{(q_s)}=0.245.$

\subsection{TFFs without gluons}

\label{SectionTFFnoGluons}

In this subsection we present the results obtained within our approach for
the \mbox{pseudoscalar meson-$\gamma$} TFFs. In fact, we have modified the
expression on the right hand side of Eq.~(\ref{TFF.01}) by adding subleading
terms in an expansion in inverse powers of $Q^{2}$. This procedure has been
already used in Refs.~\cite{Noguera:2010fe, Noguera:2012aw, Dumm:2013zoa,
Noguera:2011fv} in order to account for contributions coming e.g.\ from
higher twist operators. Here we propose to include two terms in this
expansion. In addition, let us neglect for now the gluon contribution to the
TTFs. This is consistent with a description of mesons within the nlNJL,
which has no gluon content. We have in this way
\begin{equation}
Q^{2}\,\Fmg(Q^{2}) \ = \ \int dx~\frac{1}{2}~T_{q\bar{q}%
}\left(  x,Q^{2},\mu^{2}\right)  ~\Phimq(  x,\mu^{2})  +\frac
{c}{Q^{2}}+\frac{d}{Q^{4}}\ .
\label{TFF.ng.01}
\end{equation}
In accordance with our approximation of neglecting gluon contributions, we
will use octet evolution for the whole quark DAs $\Phimq(x,\mu^{2})$.

\begin{figure}
[ptb]
\begin{center}
\includegraphics[%height=7.7145cm,
width=10cm]{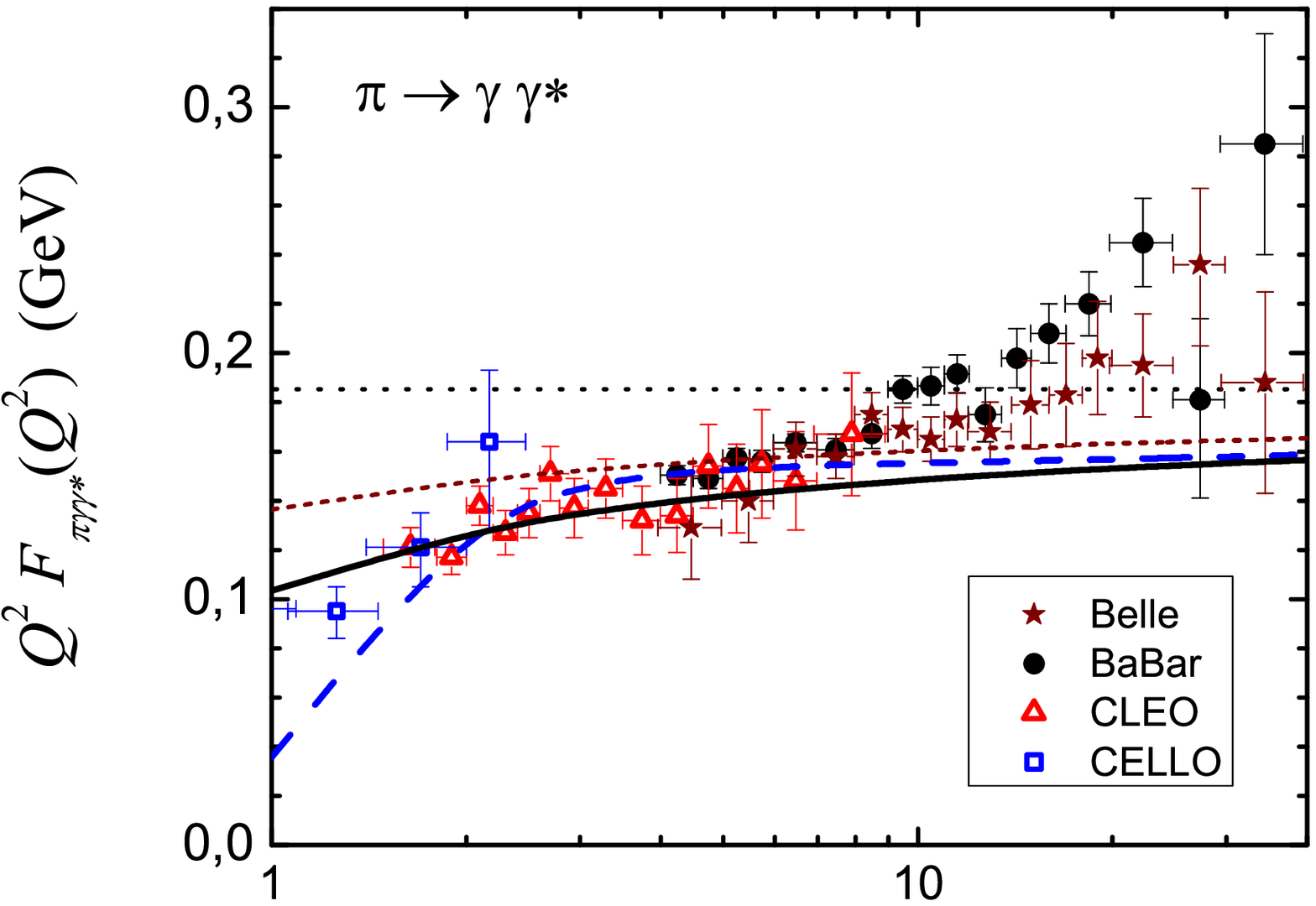}\\
\vspace{-1cm}
\includegraphics[%height=7.1127cm,
width=10cm]{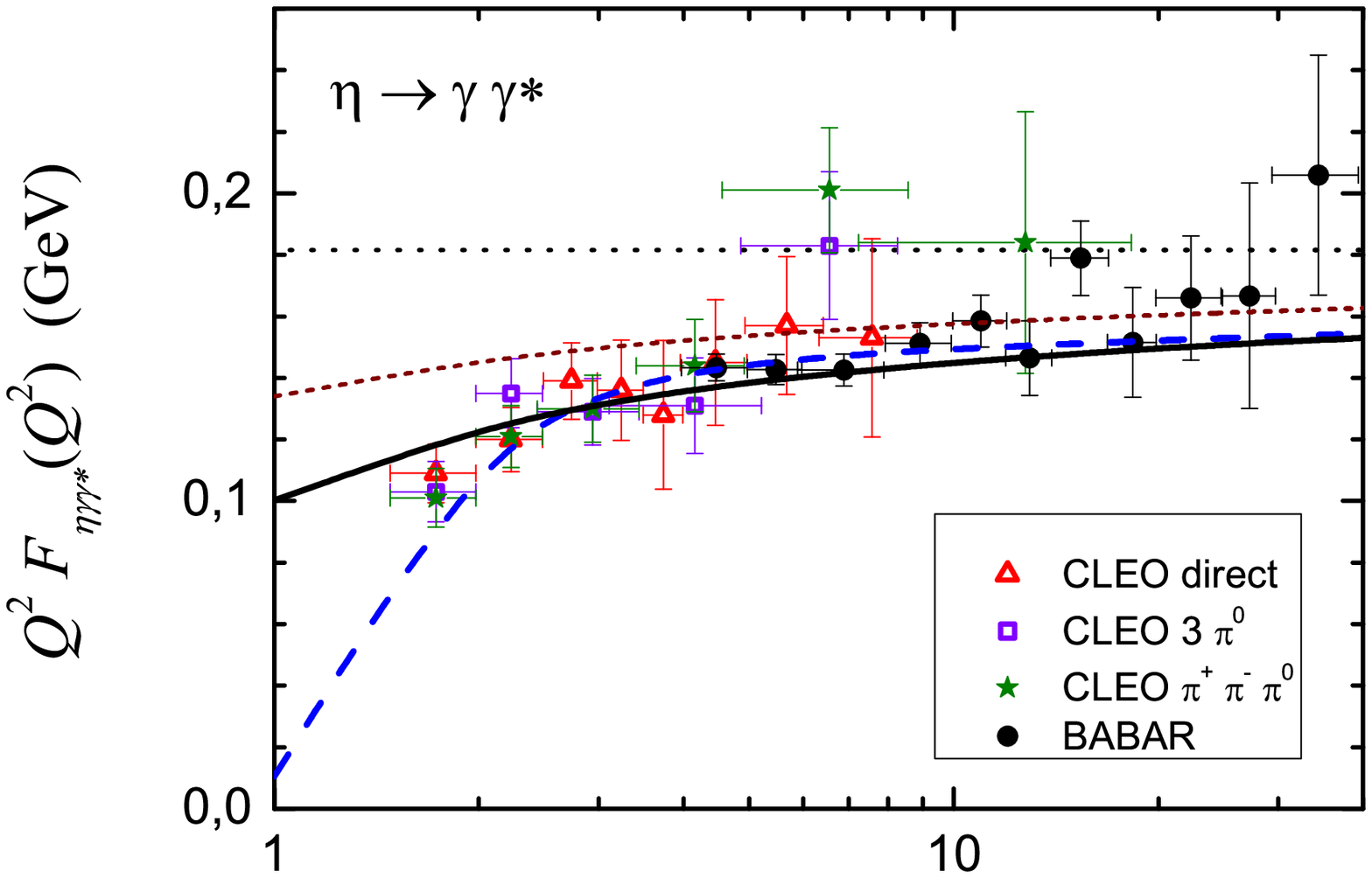}\\
\vspace{-1.2cm}
\includegraphics[%height=7.1127cm,
width=10cm]{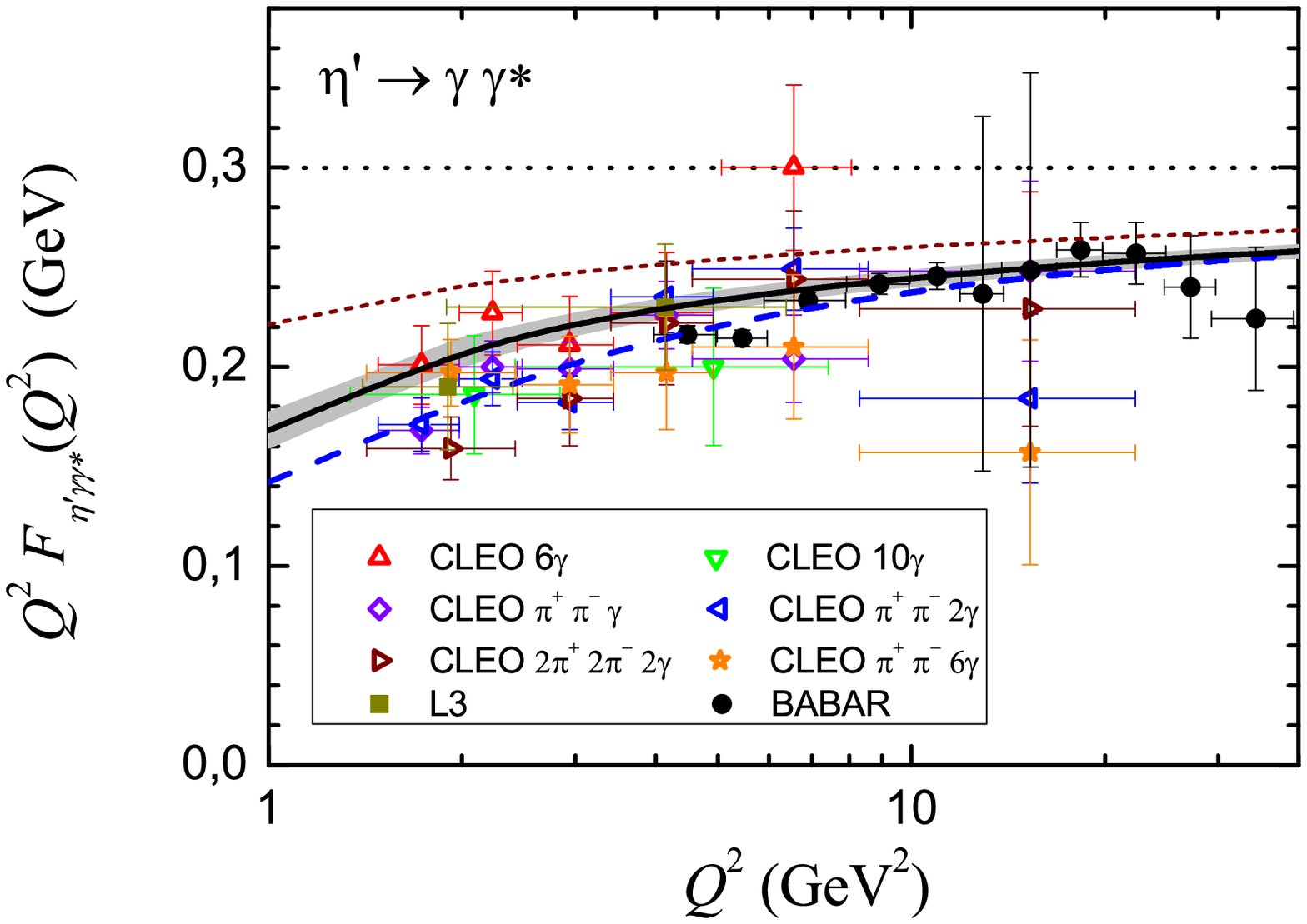}%
\caption{Theoretical $\pi$-$\gamma$, $\eta$-$\gamma$ and
$\eta^\prime$-$\gamma$ transition form factors and experimental results from
CELLO, CLEO, BaBar, Belle and L3 Collaborations. Solid lines correspond to
the case $c=d=0$, while dashed lines correspond to the values of $c$ and $d$
in Table~\ref{Table.C.D.parameters}. Short-dashed lines show the NLO
asymptotic QCD behavior (see text), and dotted lines indicate the QCD
asymptotic limits. In the case of the $\eta^\prime$-$\gamma$ TFF, the gray
region corresponds to a change in $a_{\eta^{\prime}n}^{(q_i)}$ coefficients
within a range of 15\%. Notice that in all graphs we have used a logarithmic
scale for $Q^{2}$.}
\label{FigTFFpionNGlog}%
\end{center}
\end{figure}

Our results for the $M$-$\gamma$ TFFs, where $M=\pi$, $\eta$ and
$\eta^{\prime}$, are shown in Fig.~\ref{FigTFFpionNGlog}. The curves have
been obtained by calculating the corresponding DAs at NLO, using the octet
evolution given by Eq.~(\ref{anNLO.01}). In all cases solid lines correspond
to the evaluation of the TFFs under the assumption of no higher twist
corrections, i.e.\ taking $c=d=0$, while dashed lines are obtained from
Eq.~(\ref{TFF.ng.01}) by fitting $c$ and $d$ to the experimental data. In
the case of the $\pi$-$\gamma$ and $\eta$-$\gamma$ TFFs we have considered
all world data, i.e.\ those obtained by CELLO, CLEO, BaBar and Belle
Collaborations for the $\pi$-$\gamma$ TFF and those from CLEO and BaBar for
the $\eta$-$\gamma$ TFF. On the other hand, for the $\eta^{\prime}$-$\gamma$
TFF we have considered only the data from CLEO and BaBar, in view of the
large errors in the determination of $Q^2$ values shown by L3 results (which
are also included in the figure). The dotted lines in the graphs correspond
to the LO asymptotic $Q^2\to\infty$ limits for the TFFs in QCD, namely
\begin{equation}
Q^{2}\Fmg^{\text{AsymLO}}(Q^{2})\ = \ \left\{
\begin{array}{ll}
\sqrt{2}f_{\pi} & M=\pi\\
(\sqrt{2}f_{M}^{8}+4f_{M}^{0})/\sqrt{3}=(5\sqrt{2}\fml+2f_{M}^{\,s})/3\ \ \
& M=\eta,\eta^{\prime}
\end{array}
\right.  \ \ .
\end{equation}
Finally, the short-dashed curves correspond to what we call the ``asymptotic
behavior'', obtained from Eq.~(\ref{TFF.ng.01}) by taking $c=d=0$, the
parton level amplitudes $T_{q\bar q}$ at the NLO, and the asymptotic form of
the DAs, $\Phimq(x)=\phi_{\rm asym}(x)=6x(1-x)$. One
has~\cite{delAguila:1981nk,Lepage:1980fj}
\begin{equation}
Q^{2}\Fmg^{\text{AsymNLO}}(Q^{2})  \ =\ \left(
1-5\frac{\alpha_{s}(Q^{2})}{3\,\pi}\right)  \,\left[
Q^{2}\Fmg^{\text{AsymLO}}(Q^{2})  \right] \ .
\end{equation}

In Table~\ref{Table.C.D.parameters} we quote the values of $c$ and $d$
arising from our fits, together with the number of experimental data
considered in each case and the corresponding $\chi^{2}$ values. For
comparison we also include the $\chi^{2}$ obtained when we take $c=d=0$. By
looking at the $\chi^2$ values it is seen that the introduction of higher
twist corrections through the $c$ and $d$ terms leads to a significant
improvement in the theoretical description of the data for both the
$\pi$-$\gamma$ and $\eta^{\prime}$-$\gamma$ TFFs, while the improvement is
not so important in the case of the $\eta$-$\gamma$ TFF. In this
regard, notice that the better quality of the fits is basically dominated by
the description of the low virtuality region (which has less impact in the
case of the $\eta^\prime$-$\gamma$ TFF owing to the wide dispersion of the
data). In fact, by comparing the solid and dashed curves in the figure we
observe that in the case of the $\pi$-$\gamma$ and $\eta$-$\gamma$ TFFs the
differences are ruled by the behaviors at $Q^{2}\lesssim 3$~GeV$^2$, while
for the $\eta^\prime$-$\gamma$ TFF there is a more steady deviation which
covers a region up to $Q^{2}\sim 10$~GeV$^2$. Moreover, from
Table~\ref{Table.C.D.parameters} it is seen that the signs of $c$ and $d$
are the same for $\pi$-$\gamma$ and $\eta$-$\gamma$ TFFs, whereas they are
opposite to those obtained from the fit to $\eta^{\prime}$-$\gamma$ TFF
data. This could be related with the octet character of the $\pi$ and the
prevailingly octet character of the $\eta$, which contrast with the mostly
singlet character of the $\eta^{\prime}$.

\begin{table}[tbp] \centering
\begin{tabular}
[c]{|c|c|c|c|c|c|}\hline
 \ Meson \ &
\ $c$ (GeV$^{3}$) \ &
\ $d$ (GeV$^{5}$) \ &
\ \ \ $n$ \ \ \ &
\ $\chi^{2}/n$ \ &
\ $\chi^{2}/n$ ($c=d=0$)\
\\\hline
$\pi$ & $0.130$ & $-0.234\,$\ \ \ & $50$ & $3.9$ & $6.9$\\
$\eta$ & $0.064$ & $-0.159\,$\ \ \ & $30$ & $0.76$ & $1.3$\\
$\eta^{\prime}$ & $-0.075\,$\ \ \ & $0.049$ & $40$ & $0.88$ &
$2.5$\\
\hline
\end{tabular}
\caption{Fitted values of $c$ and $d$ for the $\pi$-, $\eta$- and
$\eta^\prime$-$\gamma$ TFFs. $n$ stands for the number of experimental data
points in each case. In the last column we quote the value of $\chi^{2}$
corresponding to the choice $c=d=0$.} \label{Table.C.D.parameters}
\end{table}

Now, while higher twist effects influence the low $Q^{2}$ region of the
TFFs, it is interesting to analyze the high virtuality region from the point
of view of QCD, comparing our results with the asymptotic QCD behavior and
the asymptotic limit of the TFFs. From the graphs in
Fig.~\ref{FigTFFpionNGlog} it is seen that in all cases the introduction of
NLO corrections to the parton level subprocess amplitudes $T_{q\bar q}$
(while keeping the asymptotic limit for the DAs) has a negative contribution
to the TFFs. In addition, it is seen that in all cases the results obtained
within the nlNJL model approximate experimental data from below.

Let us comment separately the situation for each meson. In the case of
the pion, the experimental data seem to cross the asymptotic limit at some
$Q^2$ value between $\sim 10 - 20$ GeV$^2$, hence the NLO corrections go in
the wrong direction. This is a well-known problem that we have already
discussed in the context of the two-flavor version of the nlNJL model in
Ref.~\cite{Dumm:2013zoa}. In fact, the puzzling pion data can be described
by some models based on flat DAs and some cutoff in the parton
amplitudes~\cite{Radyushkin:2009zg,Noguera:2010fe, Noguera:2012aw}.

In the case of the $\eta$-$\gamma$ TFF, even if experimental data for
$Q^{2}>10$~GeV$^2$ seem to follow the asymptotic behavior, the trend of the
data shows that it is not unlikely that the TFF crosses the QCD asymptotic
limit for higher $Q^2$~\cite{Noguera:2011fv}. In any case, according to
present experimental results, it can be said that our model provides a good
description of the TFF.

Finally, for the $\eta^\prime$-$\gamma$ TFF the experimental data lie
clearly below the asymptotic behavior, and quite far from the asymptotic QCD
limit. Once again the results obtained within the nlNJL model are shown to
be in good agreement with the data. Given the uncertainty in the numerical
calculations for the $\eta'$ DA discussed in the previous subsection, we
have studied in this case the stability of our results against some
variation in the coefficients of the Gegenbauer expansion of the quark DAs.
In order to get an estimation of the error we have considered the
$\eta^\prime$-$\gamma$ TFF for $c=d=0$, allowing for a change in
$a_{\eta^{\prime }n}^{(q_i)}$ coefficients ($n\geq2$) within a 15\% range. The
corresponding range obtained for the TFF is shown by the small gray area in
Fig.~\ref{FigTFFpionNGlog}. In general we can state that this error does not
affect qualitatively our results.

\section{The effect of gluons}

In this section we discuss the possible effect of the presence of gluon
components in the DAs. According to the discussion in Sec.\ II.C, it is
natural to carry out our analyses using the octet-singlet basis for the
quark distribution amplitudes. At any scale $\mu$, we can use
Eqs.~(\ref{DAEv.01}) to obtain the octet and singlet quark DAs from the
flavor ones, and analogous expressions can be written for the
coefficients of the Gegenbauer expansion. Let us assume that we know the
flavor DAs or, equivalently, the coefficients of the Gegenbauer
expansion at some scale $\bar \mu_0$. For the octet and singlet Gegenbauer
coefficients we have
\begin{eqnarray}
\aqq8mn (\bar \mu_0) & = &\frac{1}{\sqrt{3}\,\fm8}
\left[ \fml\,\aqlmn(\bar \mu_0)  - \sqrt{2}\,\fms
\,\aqsmn(\bar \mu_0) \right]\ , \nonumber \\
a_{Mn}^{(q_0)} (\bar \mu_0)  & = & \frac{1}{\sqrt{3}\,\fmq0}
\left[\sqrt{2}\,\fml\,\aqlmn(\bar \mu_0) +
\fms\,\aqsmn(\bar \mu_0)
\right]\ .
\label{mu0bar}
\end{eqnarray}
At LO, the evolution from $\bar\mu_0$ up to a higher scale $\mu$ is
obtained from Eqs.~(\ref{DAEv.15}) and (\ref{DAEv.08}). In particular, for
quark singlet and gluon coefficients one has
\begin{align}
\aplmn\left( \bar \mu_{0}\right)   &  =\frac{ \aqqq0mn
\left( \bar \mu_{0}\right)  -\rho_{n}^{\left(  -\right)  }\,\agmn\left(
\bar \mu_{0}\right)  }{1-\rho_{n}^{\left(  -\right)  }\,\rho_{n}^{\left(  +\right)
}}\ ,\nonumber\\
\amimn\left( \bar \mu_{0}\right)   &  =\frac{ \agmn
\left( \bar \mu_{0}\right)  -\rho_{n}^{\left(  +\right)  }\,\aqqq0mn\left(
\bar \mu_{0}\right)  }{1-\rho_{n}^{\left(  -\right)  }\,\rho_{n}^{\left(
+\right)}}\ ,
\end{align}
hence
\begin{align}
\aqqq0mn\left(  \mu\right)   &  =\frac{1}{1-\rho_{n}^{\left(  -\right)
}\rho_{n}^{\left(  +\right)  }}\left\{  \left[  \left(  \frac{\alpha
_{s}\left( \bar \mu_{0}\right)  }{\alpha_{s}\left(  \mu\right)  }\right)
^{\gamma_{n}^{\left(  +\right)  }/\beta_{0}}-\rho_{n}^{\left(  +\right)  }%
\rho_{n}^{\left(  -\right)  }\left(  \frac{\alpha_{s}\left( \bar \mu_{0}\right)
}{\alpha_{s}\left(  \mu\right)  }\right)  ^{\gamma_{n}^{\left(  -\right)
}/\beta_{0}}\right]  \,\aqqq0mn\left( \bar \mu_{0}\right)  \right. \nonumber\\
&  ~~~~~~~~~~~~~~~~~~~~~~~~\left.  -\rho_{n}^{\left(  -\right)  }\left[
\left(  \frac{\alpha_{s}\left( \bar \mu_{0}\right)  }{\alpha_{s}\left(
\mu\right)  }\right)  ^{\gamma_{n}^{\left(  +\right)  }/\beta_{0}}-\left(
\frac{\alpha_{s}\left( \bar \mu_{0}\right)  }{\alpha_{s}\left(  \mu\right)
}\right)  ^{\gamma_{n}^{\left(  -\right)  }/\beta_{0}}\right]  \,a_{Mn}^{(g)}
\left( \bar \mu_{0}\right)  \right\}\ ,  \label{DaEvG.05}
\end{align}
\begin{align}
\agmn\left(  \mu\right)   &  =\frac{1}{1-\rho_{n}^{\left(  -\right)
}\rho_{n}^{\left(  +\right)  }}\left\{  \rho_{n}^{\left(  +\right)  }\left[
\left(  \frac{\alpha_{s}\left( \bar \mu_{0}\right)  }{\alpha_{s}\left(
\mu\right)  }\right)  ^{\gamma_{n}^{\left(  +\right)  }/\beta_{0}}-\left(
\frac{\alpha_{s}\left( \bar \mu_{0}\right)  }{\alpha_{s}\left(  \mu\right)
}\right)  ^{\gamma_{n}^{\left(  -\right)  }/\beta_{0}}\right]  \,\aqqq0mn
\left( \bar \mu_{0}\right)  \right. \nonumber\\
&  ~~~~~~~~~~~~~~~~~~~~~~\left.  +\left[  \left(  \frac{\alpha_{s}\left(
\bar \mu_{0}\right)  }{\alpha_{s}\left(  \mu\right)  }\right)  ^{\gamma
_{n}^{\left(  -\right)  }/\beta_{0}}-\rho_{n}^{\left(  +\right)  }\rho
_{n}^{\left(  -\right)  }\left(  \frac{\alpha_{s}\left( \bar \mu_{0}\right)
}{\alpha_{s}\left(  \mu\right)  }\right)  ^{\gamma_{n}^{\left(  +\right)
}/\beta_{0}}\right]  \,\agmn\left( \bar \mu_{0}\right)  \right\}\ .
\label{DaEvG.06}
\end{align}

In fact, to the order we are working, we should compute the NLO evolution of
the DAs. At NLO the evolution of $\aqimn\left( \mu\right)$ coefficients for
different order $n$ become mixed, as one can see from Eq.~(\ref{anNLO.01})
for the case of octet components. However, the impact of NLO corrections to
the evolution of these coefficients is not significant in comparison with
the corresponding corrections for the subprocess amplitudes $T_{q\bar q}$
and $T_{gg}$ given in Eqs.~(\ref{TFF.06}). Indeed, the most important effect
on the DAs when going from LO to NLO evolution comes from the change in the
strong coupling constant, $\alpha_{s}\left( \mu\right)$. Thus we adopt the
following prescription: we consider the NLO corrections for
$T_{q\bar{q}}\left( x,Q^{2},\mu^{2}\right)$ and
$T_{gg}\left(x,Q^{2},\mu^{2}\right)$ [given by Eqs.~(\ref{TFF.06})] together
with Eqs.~(\ref{DAEv.15}) and (\ref{DaEvG.05}-\ref{DaEvG.06}) for the
evolution of the octet and singlet DAs, respectively. In all these equations
we take the NLO running equations for the strong coupling constant
$\alpha_{s}$, given by Eq.~(\ref{ap.alfaNLO}) in App.~B. In order to test
the validity of this prescription, let us study the case of the octet DAs.
In Tables~\ref{Table.a8n.pi}, \ref{Table.a8n.eta} and \ref{Table.a8n.etap}
we quote the values of the coefficients $a_{\pi n},$ $\aetnqq8$ and
$\aetpnqq8$, respectively, evolved from $\mu=3$~GeV to $\mu =1$~GeV at LO
[i.e., using Eqs.~(\ref{DAEv.15}) and (\ref{ap.alfaLO})], at NLO [i.e.,
using Eqs.~(\ref{anNLO.01}) and (\ref{ap.alfaNLO})], and within the above
described ``mixed'' approximation, which means to take the LO evolution
equation (\ref{DAEv.15}) for the coefficients and the NLO evolution equation
(\ref{ap.alfaNLO}) for $\alpha_s$. From the values in the tables one can
conclude that the ``mixed'' approximation can be used to estimate NLO
calculations with reasonably good accuracy.

We consider here two different ways of estimating the effect of gluons in
the DAs. Our first analysis is based on the fact that in general one assumes
that the scale at which standard quark models ---with no gluon content---
can be used to describe hadron physics lies around $\mu \sim0.5-1$~GeV.
Thus, we evolve the quark DAs obtained within the nlNJL from our input scale
$\mu_0=3$~GeV to a lower energy scale, which we choose to be
$\bar\mu_{0}=0.5$~GeV, and at this lower scale we impose the condition of no
gluons. Then, for higher values of $\mu$, we allow gluon contributions to be
generated through the evolution equations, which mix quark singlet and gluon
components of the DAs. For the second analysis, once again we proceed by
using the nlNJL quark model parametrization in order to calculate the
coefficients $a_{Mn}^{(q_i)}$ ($M=\eta,\eta^\prime$) of the Gegenbauer
expansion of the DAs at the scale $\mu_0=3$~GeV. Now, for $n\leq\bar n$,
where $\bar n$ is some chosen value of $n$, we also include nonzero gluon
coefficients $\agmn(\mu_0)$, and use Eqs.\ (\ref{DaEvG.05}) and
(\ref{DaEvG.06}), with $\bar \mu_0 = \mu_0$, to evolve quark and gluon
coefficients to any other scale. The values of the gluon coefficients
$\agmn(\mu_0)$ are then determined from a fit to the experimental data for
the TFFs. For the remaining coefficients (those of order $n>\bar n$) we
proceed in the same way as in the first analysis. The consistency of this
approach can be tested by analyzing the stability of the results against
changes in the chosen value of $\bar n$. Notice that this second analysis
leads to the presence of gluons at low virtuality, which is compatible with
models that include a glueball component for the description of the
$\eta$-$\eta^{\prime}$ mixing~\cite{Mathieu:2009sg}.

\begin{table}[tbp] \centering
\begin{tabular}
[c]{|c|c|c|c|c|c|c|}\hline
$n$ & $2$ & $4$ & $6$ & $8$ & $10$ & $12$\\\hline
\multicolumn{1}{|l|}{$a_{\pi n}(3\,$GeV)} & $-0.0183$ & $-0.0324$ & $0.0048$ &
$-0.0090$ & $0.0049$ & $-0.0067$\\
\multicolumn{1}{|l|}{$a_{\pi n}(1\,$GeV) (LO)} & $-0.0264$ & $-0.0552$ &
$0.0091$ & $-0.0187$ & $0.0109$ & $-0.0158$\\
\multicolumn{1}{|l|}{$a_{\pi n}(1\,$GeV) (NLO)} & $-0.0225$ & $-0.0646$ &
$0.0075$ & $-0.0242$ & $0.0114$ & $-0.0205$\\
\multicolumn{1}{|l|}{$a_{\pi n}(1\,$GeV) (Mixed)} & $-0.0284$ & $-0.0614$ &
$0.0103$ & $-0.0216$ & $0.0127$ & $-0.0187$\\\hline
\end{tabular}
\caption{Coefficients $a_{\pi n}(\mu)$ obtained within the nlNJL quark model
at $\mu=3$~GeV and their evolution down to $\mu=1$~GeV at LO, at NLO and
using the mixed approximation.}
\label{Table.a8n.pi}
\end{table}
\begin{table}[tbp] \centering
\begin{tabular}
[c]{|c|c|c|c|c|c|c|}\hline
$n$ & $2$ & $4$ & $6$ & $8$ & $10$ & $12$\\\hline
\multicolumn{1}{|l|}{$\aetnqq8(3\,$GeV)} & $-0.0911$ & $-0.0444$ &
$0.0331$ & $-0.0173$ & $0.0021$ & $-0.0008$\\
\multicolumn{1}{|l|}{$\aetnqq8(1\,$GeV) (LO)} & $-0.1315$ & $-0.0758$ &
$0.0631$ & $-0.0358$ & $0.0046$ & $-0.0018$\\
\multicolumn{1}{|l|}{$\aetnqq8(1\,$GeV) (NLO)} & $-0.1357$ & $-0.0902$ &
$0.0695$ & $-0.0439$ & $0.0039$ & $-0.0033$\\
\multicolumn{1}{|l|}{$\aetnqq8(1\,$GeV) (Mixed)} & $-0.1413$ & $-0.0842$ &
$0.0716$ & $-0.0414$ & $0.0054$ & $-0.0022$\\\hline
\end{tabular}
\caption{Coefficients $\aetnqq8(\mu)$ obtained within the nlNJL quark model
at $\mu=3$~GeV and their evolution down to $\mu=1$~GeV at LO, at NLO and
using the mixed approximation.}
\label{Table.a8n.eta}
\end{table}
\begin{table}[tbp] \centering
\begin{tabular}
[c]{|c|c|c|c|c|c|c|}\hline
$n$ & $2$ & $4$ & $6$ & $8$ & $10$ & $12$\\\hline
\multicolumn{1}{|l|}{$\aetpnqq8(3\,$GeV)} & $-0.4317$ & $-0.0305$
& $0.1791$ & $-0.0759$ & $-0.0217$ & $0.0301$\\
\multicolumn{1}{|l|}{$\aetpnqq8(1\,$GeV) (LO)} & $-0.6232$ &
$-0.0520$ & $0.3418$ & $-0.1577$ & $-0.0483$ & $0.0708$\\
\multicolumn{1}{|l|}{$\aetpnqq8(1\,$GeV) (NLO)} & $-0.6661$ &
$-0.0733$ & $0.3930$ & $-0.1839$ & $-0.0601$ & $0.0854$\\
\multicolumn{1}{|l|}{$\aetpnqq8(1\,$GeV) (Mixed)} & $-0.6698$ &
$-0.0577$ & $0.3881$ & $-0.1821$ & $-0.0566$ & $0.0837$\\\hline
\end{tabular}
\caption{Coefficients $\aetpnqq8(\mu)$ obtained within the nlNJL quark model
at $\mu=3$~GeV and their evolution down to $\mu=1$~GeV at LO, at NLO and
using the mixed approximation.}
\label{Table.a8n.etap}
\end{table}

\subsection{First analysis}

Let us analyze the numerical results obtained for the effect of gluon
contributions according to the first analysis proposed above. As stated, we
take into account the fact that usually quark models do not include gluons
at their scale of validity $\bar \mu_{0}$, therefore we can obtain the
coefficients $\aqqq0mn$ and $\agmn$ at any scale $\mu$ from
Eqs.~(\ref{DaEvG.05}-\ref{DaEvG.06}) by imposing $\agmn( \bar \mu_{0})=0$.
Moreover, from Tables~\ref{Table.gammas} and \ref{Table.RhoMm} it is seen
that the values of the $\rho_{n}^{(+)}\rho_{n}^{(-)}$ coefficients are small
and the $\gamma_{n}^{(+)}$ anomalous dimensions are close to $\gamma^{qq}.$
Hence we can assume that the mixing with gluons will have small influence on
the singlet coefficients $\aqqq0mn(\mu)$. On the other hand, since the
values of $\rho_{n}^{(+)}$ for low $n$ are not negligible, the first gluon
coefficients $\agmn(\mu)$ of the Gegenbauer expansion could give some
appreciable contribution to $\eta$ and $\eta'$ DAs.

As discussed in Sec.~II.B, we input the shape of quark propagators at the
scale $\mu_0=3$~GeV from lattice QCD calculations. In order to connect the
DAs at this scale to those at the lower scale $\bar \mu_0$ that we use as
starting point for the QCD evolution we need some approximation. We use here
octet evolution, i.e., we begin by considering the calculated DAs at
$\mu_0=3$~GeV shown in Fig.~\ref{FigDAs}, and evolve them down to
$\bar\mu_{0}=0.5$~GeV assuming no gluon components. Then, starting from the
$\bar\mu_{0}$ scale we use the evolution equations
(\ref{DaEvG.05}-\ref{DaEvG.06}) to obtain the singlet quark and gluon DAs
(the latter, generated by the mixing in the evolution) at any $\mu$. Thus at
the scale $\bar\mu_0=0.5$~GeV we have
\begin{align}
\agmn\left(  0.5{\rm~GeV}\right)   &  = 0 \ , \nonumber \\ %\label{DaEvG.10}\\
\aqqq0mn\left(  0.5{\rm~GeV}\right)   &  = \abqqq0mn\left(
3{\rm~GeV}\right)  \left[  \frac{\alpha_{s}\left(  3{\rm~GeV}\right)
}{\alpha_{s}\left(  0.5{\rm~GeV}\right)  }\right]^{\gamma_{n}^{qq}/
\beta_{0}}\ ,
\end{align}
where
\begin{equation}
\abqqq0mn\left(3{\rm~GeV}\right) \ = \ \frac{1}{\sqrt{3}\,f_{M}^{\,0}
}\left[  \sqrt{2}\,\fml\,\aqlmn\left(  3{\rm~GeV}\right)
+\fms\,\aqsmn\left(3{\rm~GeV}\right)\right]\ .
\label{comb3gev}
\end{equation}
The values for the first coefficients $\aqlmn\left(3{\rm~GeV}\right)$ and
$\aqsmn\left( 3{\rm~GeV}\right)$ for $M=\eta$ and $M=\eta^\prime$ are those
quoted in Tables~\ref{Table.an.eta} and \ref{Table.an.etaprime}. Notice than
when evolving back from $\mu=\bar\mu_0$ to $\mu=3$~GeV using the evolution
equations (\ref{DaEvG.05}-\ref{DaEvG.06}) in general we will obtain a result
for $\aqqq0mn(3{\rm~GeV})$ different from the input value
$\abqqq0mn(3{\rm~GeV})$. However, since the anomalous dimensions
$\gamma_{n}^{(+)}$ are close to $\gamma_{n}^{qq}$ (see
Table~\ref{Table.gammas}), one expects the differences to be small for all
$n$.

In Table~\ref{Table.a0n.agn.eta.etap} we quote the first coefficients of the
Gegenbauer expansion for the quark singlet and gluon DAs at $\mu=1$~GeV. As
expected from the values of $\rho_{n}^{(+)}$ in Table~\ref{Table.RhoMm}, it
is seen that the coefficients of the gluon DA decrease rapidly with $n$. The
small value of $\aetpnq0$ for $n=2$ arises from a cancellation in the
r.h.s.\ of Eq.~(\ref{comb3gev}), which reduces significantly the value of
$\abqqq0mn(3{\rm~GeV})$.
\begin{table}[tbp]
\centering
\begin{tabular}
[c]{|c|c|c|c|c|c|c|}\hline
$n$ & $2$ & $4$ & $6$ & $8$ & $10$ & $12$\\\hline
\multicolumn{1}{|l|}{$\aetnq0(1$ GeV)} & $0.182$ & $0.107$ & $-0.268$ &
$0.094$ & $0.022$ & $-0.066$\\
\multicolumn{1}{|l|}{$\aetng(1$ GeV)} & $0.342$ & $0.062$ & $-0.072$ &
$0.014$ & $0.002$ & $-0.004$\\\hline
\multicolumn{1}{|l|}{$\aetpnq0(1$ GeV)} & $-0.022$ & $-0.128$ &
$0.036$ & $-0.039$ & $0.027$ & $0.007$\\
\multicolumn{1}{|l|}{$\aetpng(1$ GeV)} & $-0.042$ & $-0.074$ &
$0.010$ & $-0.006$ & $0.003$ & $0.0005$\\\hline
\end{tabular}
\caption{Coefficients $\aqqq0mn(\mu)$ and $\agmn(\mu)$
($M=\eta,\eta^\prime$) evolved from $\mu _{0}=0.5$~GeV to $\mu =1$~GeV according
to our first analysis of gluon contributions.}
\label{Table.a0n.agn.eta.etap}%
\end{table}%

From our calculations we find that the effect of gluon contributions to the
TFFs within this approximation is negligible. In the case of the
$\eta$-$\gamma$ TFF, the comparison with experimental data for $c=d=0$ leads
to $\chi^{2}/n=1.33$, to be compared with the value of 1.30 obtained in
absence of gluons (see Table~\ref{Table.C.D.parameters}). The corresponding
curve differs slightly from that plotted in Fig.~\ref{FigTFFpionNGlog}
(central panel, solid line). For the $\eta^{\prime}$-$\gamma$ TFF the
influence of gluons in this approximation is also imperceptible. The
comparison with data leads to $\chi^{2}/n=2.9$, somewhat above the value of
2.5 quoted in Table~\ref{Table.C.D.parameters}, whereas the corresponding
curve lies within the uncertainty range indicated by the gray region in the
lower panel of Fig.~\ref{FigTFFpionNGlog}.

\subsection{Second analysis}

As stated above, in this second analysis we allow for the presence of
nonzero gluon coefficients $a_{\eta n}^{(g)}$, $a_{\eta^\prime n}^{(g)}$ at
a low $\mu$ scale for $n\leq\bar n$, where $\bar n$ is some chosen value of
$n$, and we determine the values of these coefficients by fitting to the
experimental data for the $\eta$-$\gamma$ and $\eta^\prime$-$\gamma$ TFFs.
For the coefficients $\aqqq0mn$ and $\agmn$, with $n>\bar n$, we proceed as
in the first analysis. We consider here the cases $\bar n = 2$ and $\bar n =
4$, comparing the corresponding numerical results to get an estimation of
the stability of the approach.

Let us first take $\bar n = 2$. In this case we take the coefficients
$\aqqq0mn$ and $\agmn$ for $n\geq4$ to be the same as those calculated in
the previous analysis, therefore the corresponding values at $\mu=1$~GeV can
be read from Table \ref{Table.a0n.agn.eta.etap}. For the first Gegenbauer
coefficients $a_{\eta2}^{(q_0)}$ and $a_{\eta^\prime 2}^{(q_0)}$, at the
scale $\mu_0 = 3$~GeV we use the input provided by Eq.~(\ref{mu0bar}) with
$\bar\mu_0 = \mu_0$, taking the values of $a_{M2}^{(q_\ell)}(3{\rm~GeV})$
and $a_{M2}^{(q_s)}(3{\rm~GeV})$ for $M=\eta,\eta^\prime$ from
Tables~\ref{Table.an.eta} and \ref{Table.an.etaprime}. On the other hand,
the first gluon coefficients $a_{\eta2}^{(g)}$ and $a_{\eta^\prime2}^{(g)}$
at the scale $\mu_0 = 3$~GeV are taken as free parameters to be determined
from fits to the $\eta$-$\gamma$ and $\eta^\prime$-$\gamma$ TFF experimental
data, respectively. The theoretical values for the TFFs are obtained by
evolving the coefficients $a_{M2}^{(q_0)}$ and $a_{M2}^{(g)}$ to any scale
through the above described ``mixed'' evolution approximation. Finally we
proceed in a similar way, taking now $\bar n = 4$. Namely, for $n\geq 6$ we
use the same $\aqqq0mn$ and $\agmn$ coefficients as in the first analysis,
we obtain $a_{M2}^{(q_0)}\left(3{\rm ~GeV}\right)$ and
$a_{M4}^{(q_0)}\left(3{\rm ~GeV}\right)$ from Eq.~(\ref{mu0bar}) with $\bar
\mu_0 = \mu_0$ for $n=2$ and $n=4$, respectively, and we determine
$a_{M2}^{(g)}\left(3{\rm ~GeV}\right)$ and $a_{M4}^{(g)}\left(3{\rm
~GeV}\right)$ from fits to the experimental data for the $\eta$-$\gamma$ and
$\eta^\prime$-$\gamma$ TFFs.

To discuss our results we quote not only the values of the coefficients
$a_{Mn}^{(q_0)}$ and $a_{Mn}^{(g)}$ obtained at the input scale $\mu_0=3$~GeV
but also the corresponding values after the evolution down to 1 GeV, as it
is commonly done in the literature. This is especially relevant in this
case, since the effect of gluon contributions to the TFFs is more relevant
at low virtuality, say $Q^{2}\lesssim 3$~GeV$^{2}$. Let us start by
analyzing the results for the $\eta$ meson. From the $\bar n = 2$ fit we
obtain $a_{\eta2}^{(g)}(3{\rm ~GeV})=2.66$, with $\chi^{2}/($number of
points$)=1.30$, while from the $\bar n = 4$ fit we get
$a_{\eta2}^{(g)}(3{\rm ~GeV})=-109$ and $a_{\eta4}^{(g)}(3{\rm ~GeV})=65.5$,
with $\chi^{2}/($number of points$)=0.71$. The comparison is more feasible
when we evolve the coefficients down to $\mu=1$~GeV:
\begin{equation}
\begin{array}{lll}
\;a_{\eta2}^{(g)}(1\text{ GeV})=6.37 \hspace{1cm} &
\;\,a_{\eta2}^{(q_0)}(1\text{ GeV})=0.155 \hspace{2cm} & \bar n = 2 \ \, \\
\begin{array}{l}
a_{\eta2}^{(g)}(1\text{ GeV})=-275 \\
a_{\eta4}^{(g)}(1\text{ GeV})=278
\end{array} &
\rule{0cm}{1.2cm}\left.
\begin{array}{l}
a_{\eta2}^{(q_0)}(1\text{ GeV})=1.24\\
a_{\eta4}^{(q_0)}(1\text{ GeV})=-0.937
\end{array}
\right\}
& \bar n = 4 \ .
\end{array}
\end{equation}
Taking into account the results of our first analysis (discussed in the
previous subsection), in which we obtain $\chi^{2}/($number of
points$)=1.33$, it is seen that the $\bar n = 2$ fit shows no gain of
quality in the description of the experimental data. In addition, the $\bar
n = 4$ fit leads to a strong cancellation between the $n=2$ and $n=4$ gluon
coefficients. There is no physical reason for this cancellation, therefore
we interpret this result as a spurious solution. Thus we conclude that there
is no evidence of a significant presence of gluons in the case of the $\eta$
meson.

\begin{figure}
[ptb]
\begin{center}
\includegraphics[height=7.1127cm,width=10.1572cm]{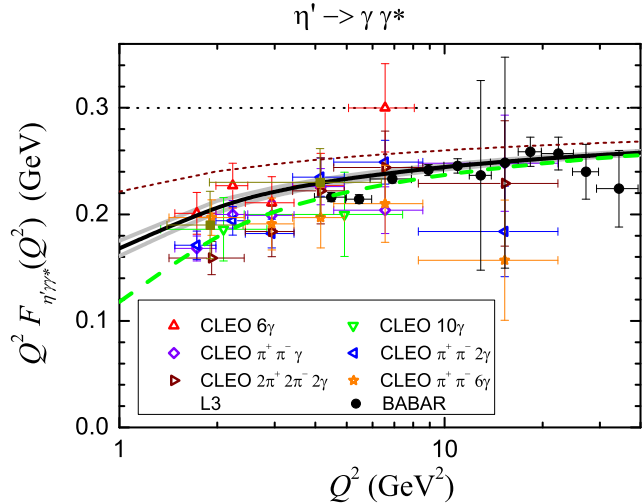}
\caption{Theoretical curves and experimental results for the
$\eta^{\prime}$-$\gamma$ TFF. The full line corresponds to the NLO result
with no gluon contributions discussed in Sec.~III, and the gray region
indicates the indetermination in the corresponding Gegenbauer coefficients.
The dashed line is the TFF obtained when the contributions of gluons are
fitted. Short-dashed and dotted lines correspond to the NLO asymptotic
behavior of the TFF (see discussion in Sec.~III) and the asymptotic limit,
respectively. Notice the usage of a logarithmic scale for $Q^{2}$.}
\label{FigTFFetapGLog}
\end{center}
\end{figure}

For the $\eta^{\prime}$ meson the $\bar n=2$ fit leads to
$a_{\eta^{\prime}2}^{(g)}(3\text{ GeV})=4.31$, while from the $\bar n=4$ fit
we obtain $a_{\eta^{\prime}2}^{(g)}(3\text{ GeV})=4.38$ and
$a_{\eta^{\prime}4}^{(g)}(3\text{ GeV})=-0.049$. The quality of the fit is
approximately the same in both cases, with $\chi^{2}/($number of
points$)=0.91.$ Evolving these coefficients to $\mu=1$~GeV we obtain
\begin{equation}
\begin{array}{lll}
\;a_{\eta^\prime 2}^{(g)}(1\text{ GeV})=10.9 \hspace{1cm} &
\;\,a_{\eta^\prime 2}^{(q_0)}(1\text{ GeV})=-0.064 \hspace{2cm} & \bar n = 2 \ \, \\
\begin{array}{l}
a_{\eta^\prime 2}^{(g)}(1\text{ GeV})=11.1 \\
a_{\eta^\prime 4}^{(g)}(1\text{ GeV})=-0.097
\end{array} &
\rule{0cm}{1.2cm}\left.
\begin{array}{l}
a_{\eta^\prime 2}^{(q_0)}(1\text{ GeV})=-0.065\\
a_{\eta^\prime 4}^{(q_0)}(1\text{ GeV})=-0.127
\end{array}
\right\}
& \bar n = 4\ .
\end{array}
\label{a0g.10}
\end{equation}
Contrary to the case of the $\eta$-$\gamma$ TFF, now we observe that there
is a significant gain of quality in the description of the experimental
values in comparison with the results from our first analysis and with those
quoted in Sec.~III. We recall that the latter, obtained under the assumption
of no gluon contributions to the $\eta^\prime$ DA, lead to a fit of
$\eta^\prime$-$\gamma$ TFF with $\chi^2/n =2.5$ (see
Table~\ref{Table.C.D.parameters}). Moreover, although the $\bar n=4$ fit has
one more free parameter with respect to the case $\bar n=2$, the theoretical
description of the $\eta^\prime$-$\gamma$ TFF is approximately the same in
both cases. Our result is shown by the dashed line in
Fig.~\ref{FigTFFetapGLog} ($\bar n=2$ and $\bar n=4$ fits are
indistinguishable). For comparison we also include in the figure the previous
NLO result with no gluon contribution (full line, indetermination indicated
by the grey band), the ``asymptotic behavior'', according to the definition
in Sec.~III (short-dashed line), and the asymptotic $Q^2\to\infty$ value
(dotted line). Our analysis shows that the gluon contribution is sizable in
the case of the $\eta^{\prime}$ meson. From the figure it is seen that in
the low virtuality region the difference between our NLO calculation and the
asymptotic behavior is similar to the difference between the present fit and
the NLO result. In fact, the result obtained after considering the fitted
gluon contributions to the $\eta^\prime$ DA is comparable to that arising
from the inclusion of higher twist contributions, discussed in the previous
section.

Finally, it is interesting to compare our results with those obtained
in Refs.~\cite{Kroll:2002nt, Agaev:2003kb, Kroll:2013iwa}. The authors of
these articles perform model independent fits of the $\eta$-$\gamma$ and
$\eta^\prime$-$\gamma$ TFFs, considering only the $n=2$ coefficients of the
Gegenbauer expansions of the DAs. Moreover, they assume meson independence
of the quark and gluon DAs, i.e.~they take
$\phi_{\eta}^{(q_i)}=\phi_{\eta^{\prime}}^{(q_i)}$ and
$\phi_{\eta}^{(g)}=\phi_{\eta^{\prime}}^{(g)}$. In this way they end up with
only three free parameters, namely the coefficients
$a_{\eta^{(\prime)}2}^{(q_8)}$, $a_{\eta^{(\prime)}2}^{(q_0)}$ and
$a_{\eta^{(\prime)}2}^{(g)}$. The analyses carried out in those papers,
considering various fits under different conditions, show that the results
are quite stable within the quoted errors. Let us take here as
representative values the default results from Ref.~\cite{Kroll:2013iwa},
namely $a_{\eta^{(\prime)}2}^{(q_8)}(1\text{~GeV})=-0.05\pm0.02$,
$a_{\eta^{(\prime)}2}^{(q_0)}(1\text{~GeV})=-0.12\pm0.01$ and
$a_{\eta^{(\prime)}2}^{(g)}(1\text{~GeV})=19\pm5$, as well as the results in
Eq.~(63) of Ref.~\cite{Agaev:2003kb}, which translated to our notation lead
to $a_{\eta^{(\prime)}2}^{(q_0)}(1\text{~GeV})=-0.12\pm0.11$ and
$a_{\eta^{(\prime)}2}^{(g)}(1\text{~GeV})=18.2\pm 4.5$. It is worth noticing
that our results do not support the hypothesis of meson independence of
quark and gluon DAs, in fact, we find significative differences between
them. Nevertheless, it is seen that the values obtained from our analysis
are consistent with the above results. Indeed, considering
Eq.~(\ref{DAq.01}), and taking values of meson decay constants from
Table~\ref{Table.pheno}, it is seen that that the coefficients of
$\phi_{M}^{(q_8)}$ are basically determined by the $\eta$-$\gamma$ TFF,
whereas those of $\phi_{M}^{(q_0)}$ (and also $\phi_{M}^{(g)}$) are mainly
fixed by the $\eta^{\prime}$-$\gamma$ TFF. Therefore, the value of
$a_{\eta^{(\prime)}2}^{(q_8)}$ in Ref.~\cite{Kroll:2013iwa} should be
compared with our result in Table~\ref{Table.a8n.eta},
$a_{\eta2}^{(q_8)}=-0.14$, while the results for
$a_{\eta^{(\prime)}2}^{(q_0)}$ and $a_{\eta^{(\prime)}2}^{(g)}$ in
Ref.~\cite{Kroll:2013iwa} and Ref.~\cite{Agaev:2003kb} are to be compared
with our values $a_{\eta^{\prime}2}^{(q_0)}=-0.06$ and
$a_{\eta^{\prime}2}^{(g)}=11$, see Eq.~(\ref{a0g.10}). Taking into account
the theoretical and experimental uncertainties, we conclude that the values
quoted in Refs.~\cite{Kroll:2013iwa} and~\cite{Agaev:2003kb} are compatible
with each other and with our results.

\section{Conclusions}

In this work we have evaluated the quark DAs for the $\eta$ and
$\eta^{\prime}$ mesons and the associated $\eta$-$\gamma$ and
$\eta^{\prime}$-$\gamma$ TFFs within the framework of a nonlocal
Nambu--Jona-Lasinio model. This approach, which has been shown to provide a
successful description of various meson observables~\cite{Noguera:2008,
Carlomagno:2013ona}, has been previously considered in
Ref.~\cite{Dumm:2013zoa} for the study of the $\pi$ meson DA and the
$\pi$-$\gamma$ TFF. Since the theoretical framework satisfies all basic
symmetry requirements (i.e.~chiral, Poincar\'{e} and local electromagnetic
gauge invariances), the quark DAs turn out to be naturally normalized within
this scheme.

One of the main ingredients of our model is the quark propagator, which by
construction shows a momentum dependence consistent with lattice QCD
results. The calculated quark DAs have to be therefore associated to the
momentum scale of lattice data, namely 3 GeV \cite{Parappilly:2005ei}. In
general, the comparison of any observable related to the quark DAs (as e.g.
the $M$-$\gamma$ TFF) with experimental data will require a perturbative
evolution of the results obtained at this reference scale. Here we have
carried out this evolution up to NLO accuracy in $\alpha_{s}$, neglecting
the mixing between Gegenbauer coefficients of different orders for the
singlet quark and gluon DAs.

From the obtained quark DAs at the scale of 3 GeV we observe the following
features: (i) whereas our $\pi$DA is not far from the asymptotic
distribution $\phi_{\text{asym}}\left(  x\right)=6x\left(  1-x\right)$, the
$\eta$ and $\eta^{\prime}$ quark DAs move away from the asymptotic behavior,
the departure being larger the larger the meson mass is; (ii) all DAs show two
maxima, and this structure arises from the nonlocal genuine contributions in
Eq.~(\ref{DA.11}); (iii) in all cases the DAs go to zero rather fast near
$x=0$ and $x=1$, supporting the idea of suppression of the kinematic end
points \cite{Mikhailov:2009kf, Mikhailov:2010ud}. Another outcome of our
results is that when the DAs are expanded in Gegenbauer polynomials we find
that the absolute values of the corresponding coefficients decrease rather
slowly with $n$, in contrast with usual assumptions.

Concerning the evaluation of the $M$-$\gamma$ TFFs, we have found that in
general NLO corrections lead to a suppression of $Q^{2}F\left(
Q^{2}\right)$. Although this represents a problem regarding the explanation
of the already challenging experimental scenario for the $\pi$-$\gamma$ TFF,
the corrections go in the right direction in the case of the $\eta$-$\gamma$
and $\eta^{\prime}$-$\gamma$ TFFs. An important difference between the case
of the $\pi$ meson and those of the $\eta$ and $\eta^{\prime}$ mesons is
that $\eta$ and $\eta'$ states can include a gluon-gluon component. In this
regard, we have firstly performed an analysis in which these gluon
components have been neglected for all $Q^{2}$ values, while higher twist
corrections have been taken into account by adding $1/Q^{2}$ and $1/Q^{4}$
terms to the dominant twist 2 contribution provided by the DAs. Then we have
fitted these contributions to $M$-$\gamma$ TFF data. From our results it is
seen that the effect of higher twist corrections is more important for the
$\pi$-$\gamma$ and $\eta$-$\gamma$ TFFs, particularly for
$Q^{2}\lesssim3$~GeV$^{2}$. Moreover, it is seen that the signs of the
corresponding contributions are the same in both cases. Conversely, for the
$\eta^{\prime}$-$\gamma$ TFF, contrary to what it should be expected, the
higher twist corrections appear to be less concentrated in the low
virtuality region.

Finally, we have investigated the effect of two-gluon components of the
$\eta$ and $\eta^{\prime}$ mesons to leading-twist accuracy, considering NLO
perturbative QCD and neglecting the mixing between Gegenbauer coefficients
of different orders. From the numerical analysis it is found that the
evolution equations do not generate an appreciable contribution if we assume
that meson DAs include no gluons at low virtuality. On the other hand, if we
allow for the presence of gluon-gluon components in the $\eta$ and
$\eta^\prime$ DAs at low momentum scales, it is seen that the experimental
data for the corresponding TFFs suggest an important gluon component in the
$\eta^{\prime}$ state and a less important one in the $\eta$ state.
According to the discussion in Sec.~IV.B, our results for the first
Gegenbauer coefficients of quark DAs at the scale of 1 GeV are
\begin{equation*}
\begin{array}{rclcrcl}
a_{\eta2}^{\left(  q_{8}\right)  }\left(  1\text{~GeV}\right)   &
= & -0.14 & \hspace{2cm} & a_{\eta4}^{\left(q_{8}\right)}
\left(1\text{~GeV}\right) & = & -0.08 \\
a_{\eta2}^{\left(  q_{0}\right)  }\left(  1\text{~GeV}\right)   &
= & 0.18 & & a_{\eta4}^{\left(  q_{0}\right)  }
\left(  1\text{~GeV}\right) & = & 0.11
\end{array}
\end{equation*}
\begin{equation*}
\begin{array}{rclcrcl}
a_{\eta^{\prime}2}^{\left(  q_{8}\right)  }\left(  1\text{~GeV}\right)  &
= & -0.67 & \hspace{2cm} & a_{\eta^{\prime}4}^{\left(  q_{8}\right)  }
\left( 1\text{~GeV}\right) & = & -0.06\\
a_{\eta^{\prime}2}^{\left(  q_{0}\right)  }\left(  1\text{GeV}\right)   &
= & -0.06 & & a_{\eta^{\prime}4}^{\left(  q_{0}\right)  }
\left(1\text{~GeV}\right) & = & -0.13
\end{array}
\end{equation*}
For the gluon DAs, our results in the case of the $\eta$ meson are not
conclusive, whereas for the $\eta^{\prime}$ we obtain
\begin{equation*}
\begin{array}{rclcrcl}
a_{\eta^{\prime}2}^{\left(  g\right)  }\left(  1\text{~GeV}\right)
= & 11 & \hspace{2cm} & a_{\eta^{\prime}4}^{\left(  g\right)  }
\left( 1\text{~GeV}\right) & = & -0.10
\end{array}
\end{equation*}
As discussed in Sec.~IV.B, these results are found to be compatible with
previous fits for Gegenbauer coefficients quoted in
Refs.~\cite{Agaev:2003kb, Kroll:2002nt, Kroll:2013iwa}. In this way, from
our analysis we conclude that $\pi$-$\gamma$ and $\eta$-$\gamma$ TFFs are
more sensible to corrections coming from higher twist effects, while the
experimental data on the $\eta^{\prime}$-$\gamma$ TFF points to the presence
of a significant gluon-gluon component in the $\eta^{\prime}$ state.

\section*{Acknowledgments}

We thank Prof.\ S.\ Scopetta for interesting discussions. This work has been
partially funded by CONICET (Argentina) under grants No.\ PIP 578 and PIP
449, by ANPCyT (Argentina) under grants No.\ PICT-2011-0113 and
PICT-2014-0492, by the National University of La Plata (Argentina), Project
No.\ X718, by the Mineco (Spain) under contract FPA2013-47443-C2-1-P, by the
Centro de Excelencia Severo Ochoa Programme, grant SEV-2014-0398, and by
Generalitat Valenciana (Spain), grant PrometeoII/2014/066. DGD acknowledges
financial support from CONICET under the PVCE programme D2392/15.

\section*{Appendix A: Details of the model}
\label{App_NL_NJL}

In this appendix we provide some details on the calculation of the quark DAs
in Eq.~(\ref{DA.01}). We start from the Euclidean action in Eq.~(\ref{se}),
to which we add a coupling with an external axial gauge field
$\mathcal{A}^a_{\mu}$, as described in Sec.~II.B. Then we perform a standard
bosonization of the fermionic theory, introducing scalar fields
$\sigma_a(x)$, $\zeta(x)$ and pseudoscalar fields $\pi_a(x)$, together with
auxiliary fields $S_a(x)$, $P_a(x)$ and $R(x)$, with $a=0,...,8$. Details of
this procedure can be found e.g.\ in Ref.~\cite{Carlomagno:2013ona}. As in
that work, we use the stationary phase approximation, replacing the path
integral over the auxiliary fields by the corresponding argument evaluated
at the minimizing values $\tilde{S_a}(x)$, $\tilde{P_a}(x)$, and
$\tilde{R}(x)$. This leads to the equations
\begin{eqnarray}
\sigma_a(x) + G \ \tilde S_a(x) + \frac{H}{2} A_{abc} \left[ \tilde S_b(x)
\tilde S_c(x) - \tilde P_b(x) \tilde P_c(x) \right] & = & 0\ ,
\nonumber \\
\pi_a(x)    + G \, \tilde P_a(x) +  H \, A_{abc} \, \tilde S_b(x) \tilde P_c(x) & = &
0\ ,
\nonumber \\
\zeta(x)    + G \, \tilde R(x) & = & 0 \ .
\end{eqnarray}
Thus the bosonized action can be written as
\begin{eqnarray}
S_E^{\mathrm{bos}} & = & -\ln\det\mathcal{D}+
\int d^{4}x\
\bigg\{ \sigma_a(x) \tilde S_a(x) + \pi_a(x) \tilde P_a(x) +
\zeta(x) \tilde R(x) \nonumber\\
& & + \ \frac{G}{2}\ \left[ \tilde S_a(x) \tilde S_a(x) + \tilde P_a(x) \tilde P_a(x) + \tilde R(x) ^2 \right] +
\nonumber\\
& &  + \ \frac{H}{4}\ A_{abc} \left[ \tilde S_a(x) \tilde S_b(x) \tilde
S_c(x) - 3 \tilde S_a(x) \tilde P_b(x) \tilde P_c(x) \right] \bigg\}\ ,
\end{eqnarray}
where
\begin{eqnarray}
\mathcal{D}\left(  y+\frac{z}{2},y-\frac{z}{2}\right)   &  = & \gamma
_{0}\;W\left(  y+\frac{z}{2},y\right)  \gamma_{0}\
\Bigg\{\delta^{(4)}(z)\; \bigg(\!-i\rlap/\partial+m_{c}\bigg) \nonumber\\
& & + \left[  \mathcal{G}(z) \bigg[  \sigma^a \left(  y\right)
+ i\gamma_5\, \pi^a\left(  y\right)  \bigg] \lambda^a +\mathcal{F}%
(z)\ \sigma_{2}\left(  y\right)  \frac{i{\overleftrightarrow{\rlap/\partial}}%
}{2\ \varkappa_{p}}\right]  \Bigg\}\ W\left(  y,y-\frac{z}{2}\right) \ .
\label{aa}%
\end{eqnarray}

As usual, we assume that, owing to parity conservation and charge and
isospin symmetries, the fields $\sigma^a(x)$, $a=0,8$, and $\zeta(x)$ have
nontrivial translational invariant mean field values $\bar{\sigma}^{a}$ and
$\bar\zeta$, while mean field values of the remaining fields are zero. Thus
we write
\begin{eqnarray}
\label{mfaf}
\sigma_a(x) &=& \bar \sigma_a+\delta\sigma_a(x) \ , \nonumber\\
\pi_a(x) &=& \delta\pi_a(x) \ , \nonumber\\
\zeta(x) &=& \bar\zeta+\delta\zeta(x) \ .
\end{eqnarray}
Replacing in the bosonized effective action, and expanding the latter in
powers of meson fluctuations $\xi$ and powers of the gauge field
$\mathcal{A}^a_{\mu}$, we obtain
\begin{equation}
S_E^{\mathrm{bos}}=S^{({\rm MFA})}+S^{(\xi^2)}+S^{(\xi\mathcal{A})}%
+\dots\ , \label{sbos}%
\end{equation}
where only the terms relevant for our calculation have been explicitly
written.

The mean field action per unit volume reads
\begin{eqnarray}
\label{semfa}
\frac{S^{({\rm MFA})}}{V^{(4)}}
&=& 2\,N_c
\sum_{f} \int \dfrac{d^3p}{(2\pi)^3}\; \log
\left[\dfrac{Z(p)^2}{p^2 + M_f(p)^2} \right]
\nonumber \\
 & &  - \left(\bar\zeta\, \bar {R} + \dfrac{G}{2}\,\bar {R}^2 +
\dfrac{H}{4}\,\bar {S}_u\, \bar {S}_d \, \bar {S}_s \right)
- \dfrac{1}{2}\, \sum_f \left( \bar \sigma_f \bar {S}_f
+ \dfrac{G}{2}\, \bar {S}_f^2 \right)\ ,
\end{eqnarray}
where we have rotated neutral fields from the SU(3)$_F$ basis to a flavor
basis, $\sigma_a,\pi_a \to \sigma_f,\pi_f$, where $a=0,3,8$ and $f=u,d,s$,
or equivalently $f=1,2,3$. The functions $M_f(p)$ and $Z(p)$ correspond to
the momentum-dependent effective masses and WFR of quark propagators
introduced in Sec.~II.B [see Eqs.~(\ref{quarkp}) and (\ref{mz})], while
$\bar S_f$ and $\bar R$ stand for the values of the fields $\tilde{S}_f(x)$
and $\tilde{R}(x)$ within the MFA, respectively. The minimization of
$S^{({\rm MFA})}$ with respect to $\bar{\sigma}_f$ and
$\bar\zeta$ leads to the corresponding Schwinger-Dyson
equations~\cite{Carlomagno:2013ona}.

The piece of the bosonic Euclidean action that is quadratic in the meson
fluctuations can be written as
\begin{equation}
\label{spiketa}
S_E^{(\xi^2)} \ =\ \dfrac{1}{2} \int \frac{d^4 p}{(2\pi)^4} \sum_{M}\  r_M\
G_M(p^2)\  \xi_M(p)\, \bar\xi_M(-p) \ ,
\end{equation}
where meson fluctuations $\delta\sigma_a$, $\delta\pi_a$  have been
translated to a charge basis $\xi_M$, $M$ being the scalar and pseudoscalar
mesons in the lowest mass nonets ($\sigma,\pi^0$, etc.), plus the $\zeta$
field. The coefficient $r_M$ is 1 for charge eigenstates $M={\rm
a}_0^0,\sigma,f_0,\zeta,\pi^0,\eta,\eta^\prime$, and 2 for $M={\rm
a}_0^+,K_0^{\ast +},K_0^{\ast 0},\pi^+,K^+,K^0$. The full expressions for
the one-loop functions $G_M(q)$, as well as those for the above mentioned
Schwinger-Dyson equations, can be found in Ref.~\cite{Carlomagno:2013ona}.
Meson masses can be obtained by solving the equations
\begin{equation}
G_M(-m_M^2)\ =\ 0 \ .
\end{equation}
In order to obtain physical states $\tilde{\xi}_M$ one still has to
introduce a wave function renormalization factor,
\begin{equation}
\tilde{\xi}_M(p)=Z_M^{-1/2}\ \xi_M(p)\ ,
\end{equation}
where
\begin{equation}
Z_M^{-1} \ = \ \frac{dG_M(p)}{dp^2}\bigg\vert_{p^2=-m_M^2} \ = \ g_{Mqq}^{-2}.
\label{gmqq}
\end{equation}

Finally, the bilinear piece in $\xi_M$ and $\mathcal{A}^a_{\mu}$ fields in
Eq.~(\ref{sbos}) is given by
\begin{equation}
S_E^{(\xi \mathcal{A})}\ =\ \mathrm{Tr}\left[  \mathcal{D}_{0}^{-1}\;\mathcal{D}_{\xi}\;\mathcal{D}%
_{0}^{-1}\;D_{\mathcal{A}}\right]  - \;\mathrm{Tr}\left[  \mathcal{D}_{0}^{-1}\;
\mathcal{D}_{\xi \mathcal{A}}\right]  \ ,
\label{spia}
\end{equation}
where $\mathcal{D}_{\xi}$, $\mathcal{D}_{\mathcal{A}}$ and $\mathcal{D}_{\xi
\mathcal{A}}$ stand for the terms in the expansion of Eq.~(\ref{aa}) that
are linear in $\xi_M$ and/or $\mathcal{A}^a_{\mu}$. Then the meson DAs
within the nlNJL model can be obtained by taking the functional derivative
of $S^{(\xi \mathcal{A})}$ with respect to $\xi_M$ and
$\mathcal{A}^a_{\mu}$. The corresponding expressions are lengthy, and will
not be quoted here. After some work one arrives at the result in
Eqs.~(\ref{DA.01}- \ref{alphas}).

It is worth noticing that, owing to the bilocal character of the current in
Eq.~(\ref{DAq.02}), one gets an extra delta function that involves the $+$
components of the momenta. Namely, if $\Gamma$ represents some operator that
includes dirac and flavor matrices, one has
\begin{eqnarray}
\int\frac{dz^{-}}{2\pi}\left. \, \bar{\psi}\left(-\frac
{z}{2}\right)  \Gamma\,\psi\left(\frac{z}{2}\right)  \right\vert
_{z^{+}=0,\,\vec{z}_{T}=0}\,e^{iP^{+}z^{-}(x-\frac{1}{2})} & = & \\
& & \hspace{-5cm}\int\frac{d^{4}p_{1}}{(2\pi)^{4}}\frac{d^{4}p_{2}}{(2\pi)^{4}}
\ \delta\left( P^{+}\Big(x-\frac{1}{2}\Big)-\frac{p_{1}^{+}+p_{2}^{+}}{2}\right)
\,\bar{\psi}_{p_{2}}\,\Gamma\,\psi_{p_{1}}\ .
\end{eqnarray}

The numerical results for meson masses and weak decay constants obtained
within the present nonlocal model, taking the parameters in
Table~\ref{Table.param}, are listed in Table~\ref{Table.pheno}.

\begin{table} [htb]
%\label{resultados}
\begin{center}
\begin{tabular}{ccccc ccccc c}
\hline
\hline
& \hspace*{0.3cm} $m_{\pi}$ \hspace*{0.3cm} & \hspace*{0.3cm} $m_{K}$ \hspace*{0.3cm}   & \hspace*{0.3cm} $m_\eta$ \hspace*{0.3cm} &
\hspace*{0.3cm} $m_{\eta^{\prime}}$ \hspace*{0.3cm}
& \hspace*{0.1cm} $f_\pi$ \hspace*{0.1cm}  & \hspace*{0.1cm} $f_K/f_{\pi}$ \hspace*{0.1cm} & \hspace*{0.1cm} $f_{\eta}^0/f_{\pi}$ \hspace*{0.1cm} &
\hspace*{0.1cm} $f_{\eta}^8/f_{\pi}$ \hspace*{0.1cm}
& \hspace*{0.1cm} $f_{\eta^{\prime}}^0/f_{\pi}$ \hspace*{0.1cm} & \hspace*{0.1cm} $f_{\eta^{\prime}}^8/f_{\pi}$ \hspace*{0.1cm} \vspace{-.2cm}\\
%\cline{2-11}
&  (MeV)       &   (MeV)     &    (MeV)     &      (MeV)             &  (MeV)     &               &                      &                      &                                &                                \\
\hline
Model   &  139$^\ast$   &   495$^\ast$ &    523     &     958$^\ast$  &  92.4$^\ast$ &    1.17       &     0.209     &      1.085  &   1.496   &  $-$0.463\ \   \\
Empirical     &  139\ \   &   495\ \    &    547     &     958\ \              &   92.4\ \    &    1.22       &     0.187  &      1.174  &   1.155   &  $-$0.456\ \   \\
\hline
\hline
\end{tabular}
\end{center}
\caption{Numerical results from our model and empirical values for various
phenomenological quantities. Input values are indicated with an asterisk.}
\label{Table.pheno}
\end{table}

\section*{Appendix B: NLO renormalization factors for the QCD evolution of the octet DA}

\label{App_QCD_Evolution}

We quote here the expressions for the renormalization factors
$E_{n}^{\text{NLO}}$ and $d_{n}^{k}$ needed to calculate the evolution of
the coefficients $a_{Mn}(\mu)$ in Eq.~(\ref{anNLO.01}). One has
\begin{equation}
E_{n}^{\text{NLO}}(\mu,\mu_{0})=\left(  \frac{\alpha_{s}\left(  \mu
_{0}\right)  }{\alpha_{s}\left(  \mu\right)  }\right)  ^{\gamma_{n}^{qq}%
/\beta_{0}}\left[  1+\frac{\alpha_{\mathrm{s}}(\mu)-\alpha_{\mathrm{s}}%
(\mu_{0})}{8\pi}\frac{\gamma_{n}^{qq}}{\beta_{0}}\left(  \frac{\gamma
_{n}^{(1)}}{\gamma_{n}^{qq}}-\frac{\beta_{1}}{\beta_{0}}\right)  \right]
,\nonumber
\end{equation}
where $\beta_{0}\,(\beta_{1})$ and $\gamma_{n}^{qq}(\gamma_{n}^{(1)})$ are
the LO (NLO) coefficients of the QCD $\beta$-function and the anomalous
dimensions, respectively. One has $\beta_1=102-38\,n_{f}/3$, where $n_{f}$
is the number of flavors (we take here $n_{f}=4$). The values of $\beta_0$
and $\gamma_n^{qq}$ are given in Sec.~\ref{evolution}, and analytical
expressions for $\gamma_{n}^{(1)}$ can be found in
Refs.~\cite{Floratos:1977au,GonzalezArroyo:1979df}. For the evolution of
the strong coupling constant $\alpha_{s}$ at LO we use
\begin{equation}
\alpha_{s}(\mu)\ =\ \frac{4\pi}{\beta_{0}\ln(\mu^{2}/\Lambda^{2})}\ ,
\label{ap.alfaLO}%
\end{equation}
with $\Lambda=0.224$~GeV, while at NLO we take
\begin{equation}
\alpha_{s}(\mu)\ =\ \frac{4\pi}{\beta_{0}\ln(\mu^{2}/\Lambda^{2})}\left\{
1\;-\;\frac{\beta_{1}}{\beta_{0}^{2}}\;\frac{\ln\big[\ln(\mu^{2}/\Lambda
^{2})\big]}{\ln(\mu^{2}/\Lambda^{2})}\right\}\ ,
\label{ap.alfaNLO}%
\end{equation}
with $\Lambda=0.326$~GeV.

On the other hand, the off-diagonal mixing coefficients $d_{n}^{k}$ in
Eq.~(\ref{anNLO.01}) are given by
\begin{equation}
d_{n}^{k}(\mu,\mu_{0})\ =\ \frac{M_{n}^{k}}{\gamma_{n}^{qq}-\gamma_{k}%
^{qq}-2\beta_{0}}\left\{  1-\left[  \frac{\alpha_{\mathrm{s}}(\mu)}%
{\alpha_{\mathrm{s}}(\mu_{0})}\right]  ^{[\gamma_{n}^{qq}-\gamma_{k}%
^{qq}-2\beta_{0}]/2\beta_{0}}\right\}  \ .
\label{ap.b04}
\end{equation}
Here the matrix elements $M_{n}^{k}$ are defined as
\begin{eqnarray}
M_{n}^{k}  &  = & \frac{(k+1)(k+2)(2n+3)}{(n+1)(n+2)}\left[  \gamma_{n}%
^{qq}-\gamma_{k}^{qq}\right] \nonumber\\
& & \times\left\{  \frac{8C_{F}A_{n}^{k}-\gamma_{k}^{qq}-2\beta_{0}%
}{(n-k)(n+k+3)}+4C_{F}\frac{A_{n}^{k}-S_{1}(n+1))}{(k+1)(k+2)}\right\}  \ ,
\label{ap.b05}%
\end{eqnarray}
where
\begin{equation}
A_{n}^{k}=S_{1}\left(  \frac{n+k+2}{2}\right)  -S_{1}\left(  \frac{n-k-2}%
{2}\right)  +2\,S_{1}(n-k-1)-S_{1}(n+1)\ ,
\label{ap.b06}
\end{equation}
with%
\begin{equation}
S_{1}\left(  n\right)  =\sum_{j=1}^{n}\frac{1}{j}\ .
\label{ap.b07}
\end{equation}
Numerical values of the coefficients $M_{n}^{k}$ for $n\leq12$ can be found
in Ref.~\cite{Agaev:2010aq}.

\end{document}